\documentclass[prl,twocolumn,footinbib,superscriptaddress,floatfix]{revtex4-2}

\usepackage[hidelinks,colorlinks=true,linkcolor=blue,citecolor=blue,anchorcolor=blue,filecolor=blue,menucolor=blue,runcolor=blue,urlcolor=blue]{hyperref}

\usepackage{amsmath,amssymb,bbold,graphicx,epsfig,amssymb,xcolor,hyperref,latexsym,float,bm,bbm,times,adjustbox,tikz,comment,verbatim,subfigure,mathtools,physics,nicefrac}

\usepackage[utf8]{inputenc}

\makeatother

\begin{document}

\title{The anomalous Floquet Anderson insulator in a continuously driven optical lattice}
\author{Arijit Dutta}
\affiliation{Goethe-Universit\"at, Institut f\"ur Theoretische Physik, 
 60438 Frankfurt am Main, Germany}

\author{Efe Sen}
\affiliation{Goethe-Universit\"at, Institut f\"ur Theoretische Physik, 
 60438 Frankfurt am Main, Germany}
 
\author{Jun-Hui Zheng}
\affiliation{Shaanxi Key Laboratory for Theoretical Physics Frontiers, Institute of Modern Physics, Northwest University, Xi'an, 710127, China}
\affiliation{Peng Huanwu Center for Fundamental Theory, Xi'an 710127, China}

\author{Monika Aidelsburger}
\affiliation{Max-Planck-Institut f\"ur Quantenoptik, Hans-Kopfermann-Strasse 1, 85748 Garching, Germany}
\affiliation{Fakultät f\"ur Physik, Ludwig-Maximilians-Universität, Schellingstrasse 4, 80799 M\"unchen, Germany}
\affiliation{Munich Center for Quantum Science and Technology (MCQST), Schellingstraße 4, 80799 M\"unchen, Germany}

\author{Walter Hofstetter}
\affiliation{Goethe-Universit\"at, Institut f\"ur Theoretische Physik, 
 60438 Frankfurt am Main, Germany}

\date{\today}

\begin{abstract}
The anomalous Floquet Anderson insulator (AFAI) has been theoretically predicted in step-wise periodically driven models, but its stability under more general driving protocols hasn't been determined. We show that adding disorder to the anomalous Floquet topological insulator realized with a continuous driving protocol in the experiment by K. Wintersperger et. al., Nat. Phys. \textbf{16}, 1058 (2020), supports an AFAI phase, where, for a range of disorder strengths, all the time averaged bulk states become localized, while the pumped charge in a Laughlin pump setup remains quantized.
\end{abstract}

\maketitle
Periodically-driven quantum systems have led to interesting phenomena in different experimental platforms~\cite{eckardt,rudner-rev,weitenberg-rev,eckardt-col,holthaus,hofstetter} and are particularly useful in the realization of nontrivial effective equilibrium states~\cite{cooper-rev,aidelsburger-rev,aidelsburger-rev2,hofstetter,ozawa}. The realization of topological models such as the Hofstadter~\cite{miyake,aidelsburger1,aidelsburger2}, the Haldane~\cite{jotzu,weitenberg,wintersperger,braun,zhengx} and the interacting Rice-Mele model~\cite{lohse,nakajima,walter} have been reported in ultracold atom and photonic systems~\cite{rechtsman,hafezi,mittal,maczewsky,mukherjee}. Most of these employ the high driving-frequency limit, where multi-photon absorption processes are suppressed~\cite{eckardt}. However, when the driving frequency becomes comparable to the other energy scales of the driven system, novel types of steady-state phases appear, which have no counterpart in equilibrium systems~\cite{kitagawa,rudner,rudner-rev}. New features in the band structure show up due to multiple-photon processes between neighboring bands which can survive even with weak two-body interactions~\cite{tao}. The anomalous Floquet topological insulator (AFTI), with a novel bulk-boundary correspondence was first theoretically predicted~\cite{kitagawa,rudner} with a step-driving protocol, and realized in photonic systems~\cite{maczewsky,mukherjee}. The crucial aspect for stabilizing the AFTI phase is the breaking of time-reversal symmetry by circular driving, and hence, the discrete nature of the drive does not play a major role. The AFTI system was realized in an ultracold atomic hexagonal lattice, with a continuous circular driving protocol, by modulating the amplitudes of three laser beams out of phase~\cite{wintersperger,braun,zhengx}.

Adding disorder to the AFTI phase can lead to a remarkable new phase --- the anomalous Floquet Anderson insulator (AFAI), at an intermediate disorder strength which is comparable to the driving frequency. The phase is characterized by the complete localization of all bulk states together with the existence of robust edge states at all energies. This leads to quantized pumping of charge, even when all the bulk states are localized, which is impossible in equilibrium systems. In spite of theoretical predictions in idealized models~\cite{titum,kundu}, it has not been experimentally realized, yet. A significant achievement would be to realize this phase in ultracold atoms which will, additionally, allow us to study the interplay with two-body interactions in a controlled way. 

\begin{figure}[t]
	\centering
	\includegraphics[trim={0cm 0cm 0cm 0cm}, clip=True,width=0.95\linewidth]{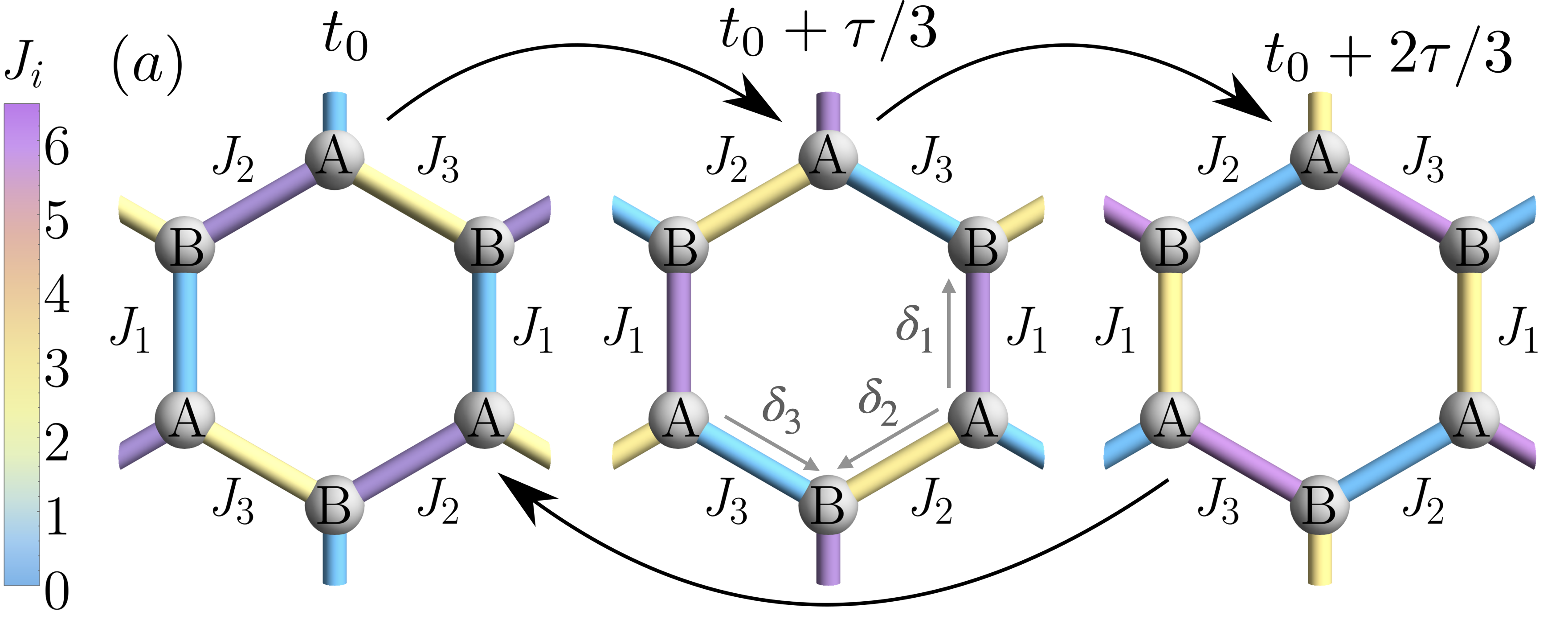}
	\includegraphics[width=\linewidth]{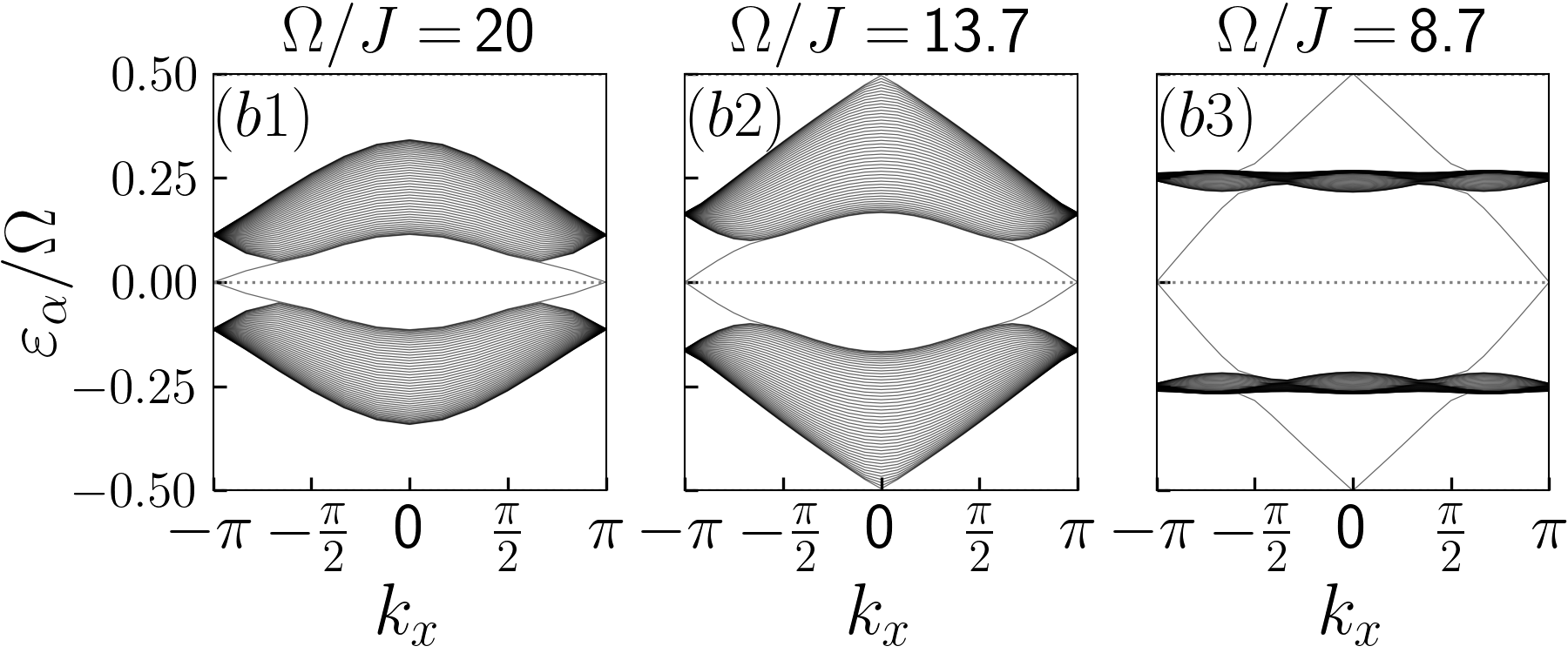}
	\caption{$(a)$~The driving protocol on the honeycomb lattice, where the hopping amplitudes at each step are represented by false colour. Quasienergy spectrum at characterisic $\Omega/J$ values, $(b1)~20$, $(b2)~13.7$ and $(b3)~8.7$ for a zigzag semi-infinite strip with $48$ unit cells and $F=2$. $(b1)$ denotes a CI regime where edge states corresponding to nonvanishing bulk Chern number appear in the $0$-gap. As the driving frequency is lowered the $\pi$-gap closes $(b2)$ and reopens, leading to an AFTI regime $(b3)$ having edge states in all gaps but vanishing Chern number of the bulk bands.}
	\label{fig:clean}
\end{figure}

Our work provides numerical evidence that it is indeed possible to stabilize the AFAI phase in the experimentally accessible parameter regimes. However, its indicators are strongly system size dependent. This is because complete localization of bulk states for 2d systems can only be realized for very large system sizes. By considering a Laughlin pump setup we show that the pumped charge over one period of the threaded flux remains quantized even when \emph{all} bulk states become localized. We work with the continuous driving protocol implemented on a honeycomb lattice as realized in Refs.~\cite{wintersperger,braun}, and add onsite disorder to it.  The honeycomb lattice has a 2-sublattice structure which we denote by labels $A$ and $B$ [Fig.\,\ref{fig:clean}(a)]. The real-space Hamiltonian is ($\hbar=1$)
\begin{align}
	H(t) &= \sum_{\bm{i}}\sum_{\gamma = 1}^{3}\left(J_{\gamma}(t)c^{\dagger}_{\bm{i}} c^{\vphantom{\dagger}}_{\bm{i}+\bm{\beta}_{\bm{i}\gamma}} + h.c.\right) + \sum_{\bm{i}}V_{\bm{i}}c^{\dagger}_{\bm{i}} c^{\vphantom{\dagger}}_{\bm{i}}\label{eq:Ht}
\end{align}
where $c_{\bm{i}}^{\dagger} (c^{\vphantom{\dagger}}_{\bm{i}})$ creates (annihilates) a spinless fermion at site $\bm{i}$, $J_{\gamma}(t) = J\exp[F\cos(\Omega t + \phi_{\gamma})]$, with $\phi_{\gamma} = \frac{2\pi}{3}(\gamma-1)$, are the hoppings across three nearest-neighbour bonds $\bm{\beta}_{\bm{i}\gamma}$ at each site $\bm{i}$. If the vector $\bm{i}$ points to a site in the $A\,(B)$-sublattice then $\bm{\beta}_{\bm{i}\gamma}=+(-)\bm{\delta}_{\gamma}$ (for $\gamma = 1,2,3$), where $\bm{\delta}_1\equiv\qty(0,a)$, $\bm{\delta}_2\equiv\qty(-\sqrt{3}a/2,-a/2)$, $\bm{\delta}_3\equiv\qty(\sqrt{3}a/2,-a/2)$ [Fig.\,\ref{fig:clean}$(a)$], $a$ is the lattice constant, $J$ is the bare hopping amplitude, $\Omega$ is the driving frequency and $F$ is a dimensionless parameter which controls the width of the bulk bands. Henceforth, we set $a=1$, $J=1$ and $F = 2$. $V_{\bm{i}}$ is an onsite disorder potential which is sampled from a uniform distribution of width $W$ and zero mean.

According to Floquet's theorem, such a time-periodic Hamiltonian admits stationary solutions, called \emph{Floquet states}, of the form $\ket{\psi_{\alpha}(t)} = \exp\qty(-i\varepsilon_{\alpha}t)\ket{u_{\alpha}(t)}$, where $\varepsilon_{\alpha}$ is the time-independent \emph{quasienergy} and $\ket{u_{\alpha}(t)} = \ket{u_{\alpha}(t+\tau)}$ is periodic with the time-period $\tau= 2\pi/\Omega$ of the drive. Hence, $\ket{u_{\alpha}(t)}$ can be expanded in its harmonics $|u^{(n)}_{\alpha}\rangle=\int_{0}^{\tau}\qty(\mathrm{d}t/\tau)\exp(in\Omega t)\ket{u_{\alpha}(t)}$, where $n$ is an integer. The Hamiltonian in Eq.~(\ref{eq:Ht}) can be diagonalized by Fourier transformation~\cite{sm} to obtain a time-independent eigenvalue problem for the Floquet harmonics $\sum_{\bm{j},m}\tilde{H}^{(n,m)}_{\bm{ij}}u_{\bm{j}\alpha}^{(m)} = \varepsilon_{\alpha}u_{\bm{i}\alpha}^{(n)}$,
where $\tilde{H}^{(n,m)}_{\bm{ij}} = \frac{1}{\tau}\int\limits_{0}^{\tau}\mathrm{d}t\,e^{i\qty(n-m)\Omega t}H_{\bm{ij}}(t)-m\Omega\delta_{nm}\delta_{\bm{ij}}$ is the ``Floquet Hamiltonian"~\cite{rudner}, $H_{\bm{ij}}(t)$ is the representation of $H(t)$ in a site-localized basis $\{\ket{\bm{i}}\}$, and $u^{(m)}_{\bm{i}\alpha}\equiv\langle \bm{i}|u^{(m)}_{\alpha}\rangle$ is the wavefunction of the $m$-th harmonic of $\ket{u_{\alpha}(t)}$ ($m$ is an integer). Henceforth, the index $\alpha$ shall be restricted to the quasienergies in the first Floquet Brillouin zone (FBZ) $-\Omega/2\leq\varepsilon_{\alpha}<\Omega/2$~\cite{comment1}.

\textit{Anomalous Floquet topological insulator (AFTI).}$-$We first consider a clean system. In Fig.~\ref{fig:clean}, we plot the dispersion of a semi-infinite strip with zigzag edges for $\Omega/J =20,\,13.7,\,8.7$, and $|m|, |n| \leq N = 9$.  We define two gaps, the $0 (\pi)$-gap having magnitude $\Delta_{0 (\pi)}$, respectively, at the center and the edge of the Floquet Brillouin zone for bulk states. For $\Omega/J = 20$, the system is a Chern insulator (CI). On decreasing $\Omega/J$, $\Delta_{\pi}$ vanishes at $\Omega/J \approx 13.7$ and the system undergoes a transition from a CI phase to an AFTI phase, akin to that realised in the experiments~\cite{wintersperger,braun}. The dispersion for a zigzag strip when $\Omega/J$ is tuned across the transition is shown in Fig.~\ref{fig:clean}. In each FBZ, the Chern number of the upper~(lower) band $(C^{\pm})$ is given by $C^{\pm}=\mp(\mathcal{W}_0 - \mathcal{W}_{\pi})$, where $\mathcal{W}_{0(\pi)}$ is an integer topological invariant for the periodically-driven bulk system, called the winding number, which counts the number of chiral edge modes within the gap at quasienergy $0(\Omega/2)$ when the system is defined on a semi-infinite strip. This justifies how the Chern number for all the bulk bands in the anomalous phase can be zero while it hosts robust chiral edge states~\cite{kitagawa,rudner}.

\begin{figure}[b]
	\centering
    \includegraphics[trim={0cm 0cm 0cm 0cm}, clip=True, width=\linewidth]{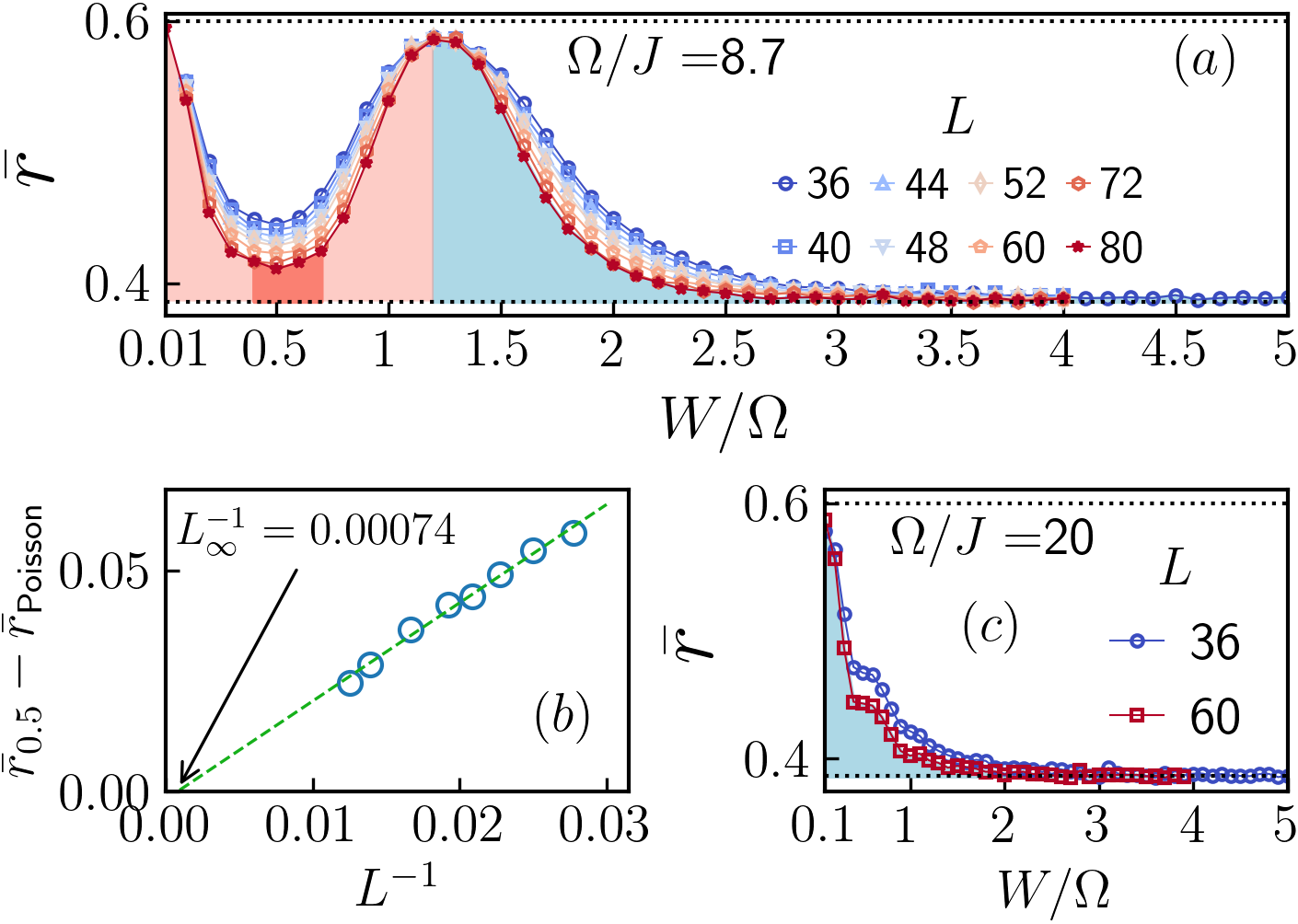}
    \caption{$(a)$ Variation of the average LSR ($\bar{r}$) with $W/\Omega$ at $\Omega/J = 8.7$. Level statistics in the delocalized regime for $W/\Omega\leq 0.01$ is characterized by $\bar{r}\approx\bar{r}_{\mbox{\scriptsize GUE}}\approx0.6$, and for the Anderson localized regime for $W/\Omega\geq 3$ by $\bar{r}\approx\bar{r}_{\mbox{\scriptsize Poisson}}\approx0.39$. $\bar{r}$ has a dip at $W/\Omega \approx 0.5$, where $\bar{r}$ approaches $\bar{r}_{\mbox{\scriptsize Poisson}}$ with increasing system size, and a peak at $W/\Omega \approx 1.2$. At the largest accessible size $(80\times80)$, the region marked in red has localized states at all quasienergies (Fig.~\ref{fig:ipr}) in the first FBZ, while the regions shown in orange and turquoise show intermediate behaviour. The red region is expected to grow while the light red and turquiose regions are expected to shrink with increasing size, and ultimately vanish in the limit of infinite system size, leading to sharp localization-delocalization-localization transitions. $(b)$ Behaviour of $\bar{r}$ at $W/\Omega=0.5$ ($\bar{r}_{0.5}$) with increasing linear dimension $L$ of the system. The best fit (green) line indicates that $\bar{r}_{0.5}$ should approach $\bar{r}_{\mbox{\scriptsize Poisson}}$ for $L\gtrsim10^3$. $(c)$~$\bar{r}$ variation with $W/\Omega$ at $\Omega/J = 20$. The system goes from the delocalized Chern insulator phase for $W/\Omega < 0.1$ to a localized Anderson insulator phase for $W/\Omega \geq 2$ with an intermediate (blue) region which shrinks with increasing size.}\label{fig:loc}
\end{figure}

\textit{Effect of disorder on the bulk.}$-$We focus on two characteristic driving frequencies, $\Omega/J = 20$ in the CI phase and $\Omega/J = 8.7$ in the AFTI phase, and study the effect of on-site disorder in the bulk. The degree of localisation in a disordered system can be characterised by the \emph{level spacing ratio} (LSR) $r_{\alpha}$ given by $r_{\alpha} = \min\{s_{\alpha}, s_{\alpha-1}\}/\max\{s_{\alpha}, s_{\alpha-1}\}$, where $\alpha$ labels the quasienergies within the first FBZ and $s_{\alpha} = \varepsilon_{\alpha+1} - \varepsilon_{\alpha}$ is the spacing between consecutive quasienergy levels. The disorder-averaged LSR distribution is given by $p(r) = \langle \sum_{\alpha}\delta(r-r_{\alpha})\rangle$, where $\langle ..\rangle$ denotes disorder averaging. Results from random matrix theory (RMT) suggest that $p(r)$ has a Poissonian form, characterized by the mean LSR ($\bar{r}$) approaching $\bar{r}_{\mbox{\scriptsize Poisson}} = 2\ln 2 -1\approx 0.39$, if all the states in the system are localized. On the other hand, $p(r)$ in a system without time-reversal invariance, in the thermodynamic limit, for extended states is given by a Wigner-Dyson form corresponding to the Gaussian unitary ensemble (GUE), which is characterized by $\bar{r}_{\mbox{\scriptsize GUE}} = 2{\sqrt{3}}/{\pi}-{1}/{2}\approx 0.60$~\cite{roux,huse,titum,guhr}. Fig.~\ref{fig:loc}$(a)$ and $(c)$ show the behaviour of $\bar{r}$ as a function of disorder strength $W/\Omega$ for $\Omega/J = 8.7$ and $20$, respectively. The two-peak structure for $\Omega/J = 8.7$, along with its size dependence, indicates the presence of a localized bulk phase around $W/\Omega \approx 0.5$, which is different from the Anderson insulator (AI) phase realized for $W/\Omega\geq3$. Moreover, the transition from this novel localized phase, which we call the anomalous localized phase, to the AI phase involves a ``critical'' point~\cite{titum}, at $W/\Omega \approx 1.2$, where $\bar{r}$ attains a maximum value. In the thermodynamic limit, we expect the transition from the anomalous localized phase to the AI phase to be ``infinitely sharp" which is supported by the finite size scaling shown in Fig.~\ref{fig:loc}(b).

\begin{figure}[t]
\centering
    \includegraphics[trim={0cm 0cm 0cm 0cm}, clip, width=\linewidth]{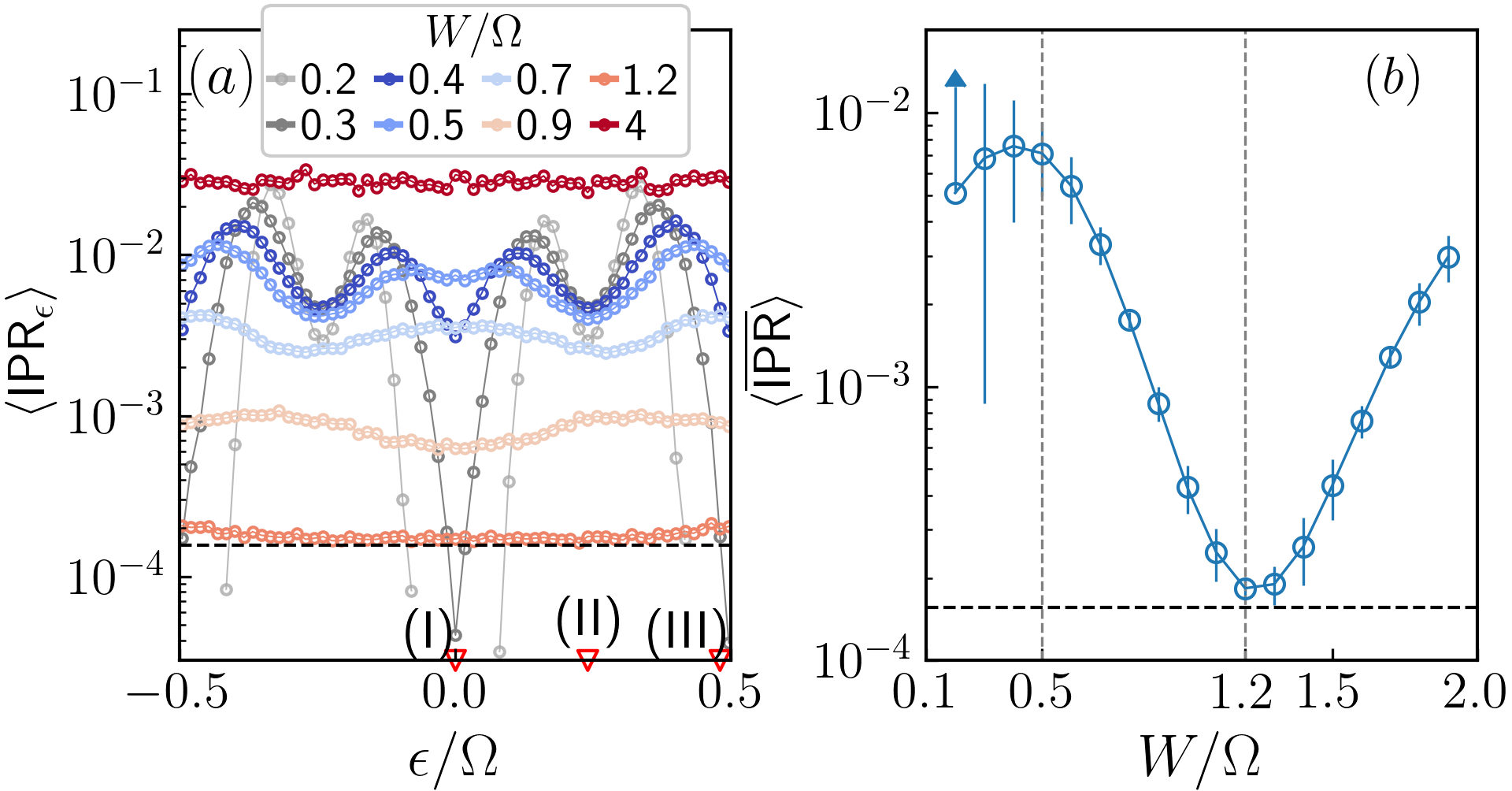}
    \includegraphics[trim={0cm 0cm 0cm 0cm}, clip, width=\linewidth]{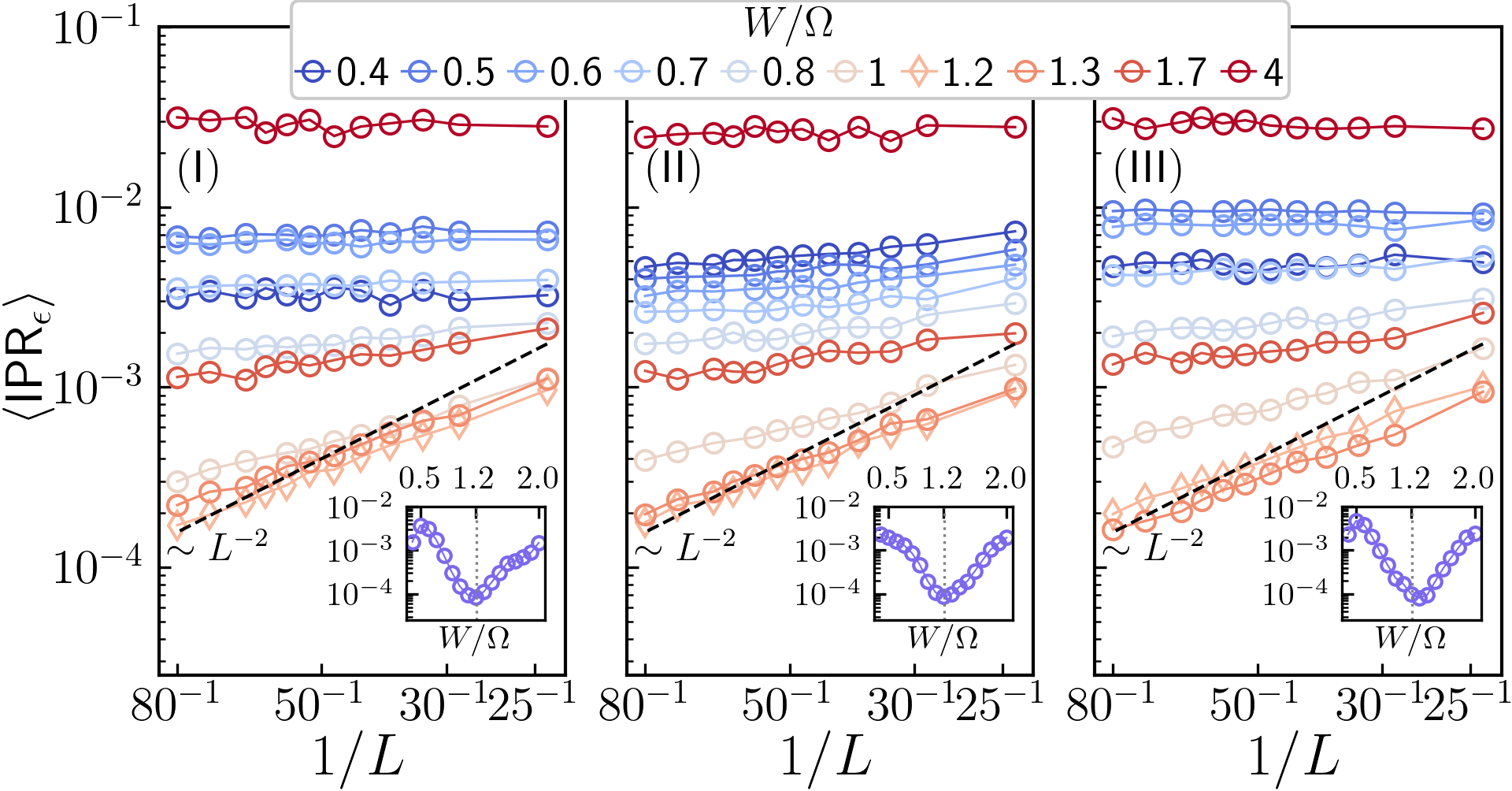}
    \caption{$(a)$ $\langle\mbox{IPR}_{\epsilon}\rangle$ for different disorder strengths $W/\Omega$ for $L=80$ at $\Omega/J=8.7$. The vertical cuts at center $(\mbox{I})$, three-quarters $(\mbox{II})$ and edge $(\mbox{III})$ of the FBZ are selected and their size-dependence is shown in the bottom panel. $(b)$ Behaviour of $\langle\overline{\mbox{IPR}}\rangle$, the $\epsilon$-average of $\langle\mbox{IPR}_{\epsilon}\rangle$, with varying $W/\Omega$. Errorbars indicate the standard deviation. For $W/\Omega = 0.2$ only one-sided error is shown since the lower bound does not fit the plot range. The horizontal dashed black lines in both $(a)$ and $(b)$ correspond to $80^{-2}$, which sets the reference level for $L^{-2}$ scaling of $\langle\mbox{IPR}_{\epsilon}\rangle$. Bottom: Log-log plots of $\langle\mbox{IPR}_{\epsilon}\rangle$ as a function of $L$ for states at the selected values of $\varepsilon/\Omega$ indicated by the red triangles in $(a)$. For $0.4 \lesssim W/\Omega\lesssim 0.7$ the system shows complete localization marked by a size independent IPR for large $L$ at all quasienergies $\epsilon$. For $W/\Omega\approx 1.2$ the system becomes delocalised for large $L$, as inferred from the slope of the corresponding trace. It should be compared with the black dashed line which corresponds to $L^{-2}$ scaling and represents the ideal delocalization limit in $2$d. The insets track the behaviour of $\langle\mbox{IPR}_{\varepsilon}\rangle$ versus $W/\Omega$ for $L = 80$. The dip occurs around $W/\Omega = 1.2$ for the three energy slices.}
    \label{fig:ipr}
\end{figure}
To investigate the localization properties of the anomalous localized phase in more detail, we consider the \emph{inverse participation ratio} (IPR) for the time-averaged state $|u_{\alpha}^{(0)}\rangle$, defined, for each disorder realization, as $\mbox{IPR}_{\epsilon} \equiv \sum_{\bm{i},\alpha}|\langle \bm{i}|u^{(0)}_{\alpha}\rangle|^{4}\delta(\epsilon-\varepsilon_{\alpha})$, where $\ket{\bm{i}}$ is the site basis state. $\mbox{IPR}_{\epsilon}$ scales as $L^{-2}$ for extended states in $2$-dimensions, where $L$ is the linear dimension. For localized states it becomes independent of system size, for sufficiently large $L$. Fig.\,\ref{fig:ipr}~(a) shows the disorder averaged IPR, $\langle\mbox{IPR}_{\epsilon}\rangle$ for $\epsilon$ in the first FBZ, for different disorder strengths $W/\Omega$. For $W/\Omega < 0.3$, the spectrum has a gap at $\epsilon/\Omega = 0$ and $\pm 0.5$, even for the largest system size which could be accessed ($80\times 80$). For $W/\Omega \approx 0.3$, the spectrum becomes gapless for large $L$, but has a gap at lower $L$ values, while for $W \geq 0.4$ the spectrum remains gapless for all the accessed $L$ values. In order to understand the overall behaviour of $\langle\mbox{IPR}_{\epsilon}\rangle$ with changing $W$, we show its mean over $\epsilon$, $\langle\overline{\mbox{IPR}}\rangle$, along with the corresponding standard deviation in Fig.\,\ref{fig:ipr}~(b). As $W/\Omega$ increases, $\langle\overline{\mbox{IPR}}\rangle$ attains its maximum value near $W/\Omega = 0.5$, indicating a maximally localized state on average, and minimum value near $W/\Omega = 1.2$, indicating that  on average, a maximum number of states are delocalized. 

To further support these observations, we show the size dependence of $\langle\mbox{IPR}_{\epsilon}\rangle$ at three characteristic quasienergies chosen at the center $(\mbox{I})$, three-quarters $(\mbox{II})$ and edge $(\mbox{III})$ of the first FBZ in the bottom panel of Fig.~\ref{fig:ipr}. These points are indicated by red triangles in Fig.\,\ref{fig:ipr} (a). We find that for $0.4 \leq W/\Omega\leq 0.7$, $\langle\mbox{IPR}_{\epsilon}\rangle$ remains $L$ independent for $L\geq 60$, across all the three $\epsilon$ slices. Hence, we expect the bulk states at all quasienergies to be localized for this range of $W$ values. Even for lower $L$ values $\langle\mbox{IPR}_{\epsilon}\rangle$ shows almost no scaling at the center and edge of the first FBZ for these values of $W/\Omega$, but shows scaling behaviour for a quasienergy in between them [slice $(\mbox{II})$]. Furthermore, at $W/\Omega\approx 1.2$ we find the emergence of $L^{-2}$ scaling for large $L$, at all the three $\epsilon$-slices, even though the disorder strength is even larger than the bandwidth of the clean system. This confirms the presence of an additional localized phase around $W/\Omega = 0.5$ and a localization-delocalization transition around $W/\Omega=1.2$, as indicated by the LSR.

\begin{figure}[t]
	\centering
 	\includegraphics[trim={0cm 0cm 0cm 0cm}, clip=True,width=0.9\linewidth]{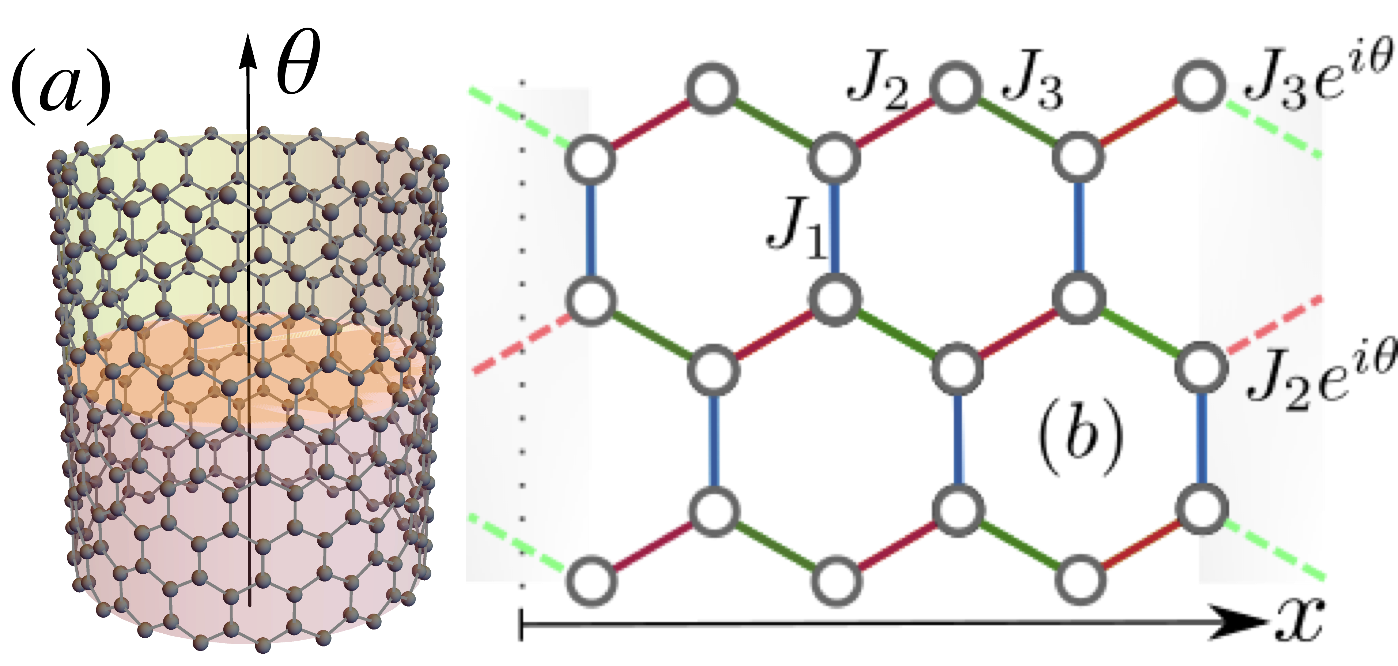}
    \includegraphics[trim={0cm 0cm 0cm 0cm}, clip=True, width=\linewidth]{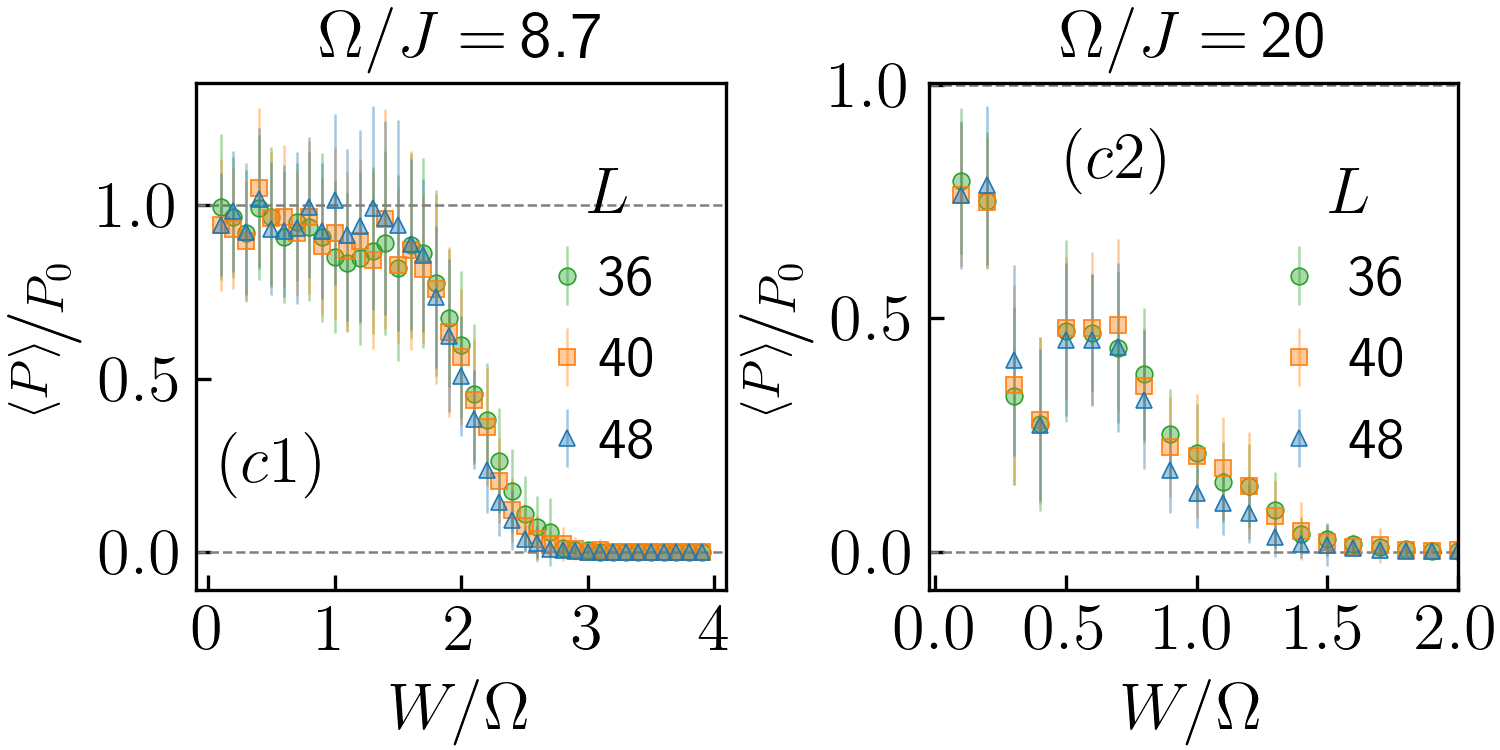}
	\caption{(a) The Laughlin pump setup where a flux $\theta$ is threaded through a cylindrical system. ($b$) The gauge choice in which additional phases $e^{i\theta}$ are acquired by the hopping amplitudes across the bonds which intersect the $x=0$ (dotted) line. Bottom: Variation of the disorder normalized average pumped charge $\langle P\rangle/P_0$, with disorder strength $W/\Omega$ for $\Omega/J = 8.7$ ($c1$) and $20$ ($c2$). For $0.1 \leq W/\Omega \lesssim 1.2$, in the anomalous phase ($c1$) $\langle P\rangle/P_0$ remains quantized to the value $1$ (within errorbars), and decays to zero with increasing disorder strength. In the CI phase ($c2$), $\langle P\rangle/P_0$ is reduced below $1$ even for the lowest disorder strength considered, and decays to zero with increasing disorder. The transitions become sharper with increasing linear dimension $L$.}\label{fig:disPC}
\end{figure}
\textit{Charge pumping.}$-$\label{sec:top}
The topological properties of the phases can by evaluated by setting up a Laughlin charge pump, where a flux $\theta$ is threaded through the system in a cylindrical geometry~\cite{laughlin,taoqin}, as shown in Fig.~\ref{fig:disPC}(a). Assuming that the zigzag edge of the cylinder is oriented along the $x$-direction, we choose a gauge such that the nearest-neighbour hopping elements across the bonds which intersect the line $x=0$ acquire an additional phase $\exp(\pm i\theta)$~\cite{titum} for hopping to the right (left) [Fig.\,\ref{fig:disPC}(b)]. The total occupancies in the upper ($\mathcal{U}$) and lower ($\mathcal{L}$) halves of the cylinder are given by $Q_{\mathcal{U}(\mathcal{L})}(\theta) = \sum_{\bm{i}\in \mathcal{U}(\mathcal{L})}Q_{\bm{i}}$, where $Q_{\bm{i}}$ is the occupancy of the site at position $\bm{i}$. The difference between the total particle numbers accumulated in the upper and lower halves of the cylinder is $\delta Q(\theta) = Q_\mathcal{U}(\theta) - Q_\mathcal{L}(\theta)$ and $P\equiv (\max\{\delta Q(\theta)\}-\min\{\delta Q(\theta)\})/2$ is the \emph{pumped charge} in one period of the threaded flux $\theta$. In a topologically nontrivial phase, the flux threading is accompanied by a discontinuity of $\delta Q$ as $\theta$ is varied between $[0,2\pi]$~\cite{taoqin}.

The site occupancies $Q_{\bm{i}}$ can be expressed in terms of the lesser Floquet Green's function $G^{<}$~\cite{taoqin,sm,neqdmft}
\begin{align}\label{eq:Qi:0}
&Q_{\bm{i}} = \sum\limits_{n=\frac{-N}{2}}^{\frac{N}{2}} 
\int\limits_{-\Omega/2}^{\Omega/2}\frac{\mathrm{d}\omega}{2\pi}\,
\lim_{\Gamma\rightarrow0^{+}} \mbox{Im}\qty[G^{<}_{\bm{i},n;\bm{i},n}(\omega;\theta)]\nonumber\\
&= \sum\limits_{\alpha}\qty(\sum\limits_{n}\abs{u^{(n)}_{\bm{i}\alpha}}^2)\qty(\sum\limits_{\bm{l},p}f(\varepsilon_{\alpha}+p\Omega)\abs{u^{(p)}_{\bm{l}\alpha}}^2)
\end{align}
where $f(\omega)\equiv\qty(1+\exp\qty(\omega/T))^{-1}$, and $n$ and $p$ are integers. In order to specify an initial state we have introduce a bath at temperature $T$ which is quadratically coupled to the system with a coupling strength $\Gamma$~\cite{sm,tao2}. However, the experimentally prepared ultracold atomic systems are essentially isolated, so we take the limit $\Gamma\rightarrow 0$ and set $T=0.01$ in Eq.~(\ref{eq:Qi:0}). We use $Q_{\bm{i}}$ to calculate the disorder-averaged steady-state pumped charge $\langle P\rangle$, normalized by its reference value $P_0$ in the clean system, by tracking the $\theta$ dependence of $\delta Q$ for every disorder realisation. $\langle P\rangle/P_0$ has been plotted in Fig.~\ref{fig:disPC} $(c1)$ for $\Omega/J=8.7$ in the anomalous localized phase and in Fig.~\ref{fig:disPC} $(c2)$ for $\Omega/J=20$ in the CI phase. We find that $\langle P\rangle/P_0$ remains quantized in the anomalous phase for $W/\Omega < 1.2$, while it decreases rapidly with increasing $W/\Omega$ in the CI phase~\cite{comment2}. This means the phase at $\Omega/J = 8.7$ supports quantized charge pumping through the edge states while its time-averaged bulk states remain completely localized for $0.4\leq W/\Omega\leq 0.7$. This is the signature of the AFAI phase, as discussed in Ref.~\cite{titum}, which supports one chiral edge mode at each edge of the cylinder, and the two edge modes have opposite chiralities.

\textit{Discussion.}$-$When all the bulk states are localized then the Chern number at any quasienergy must be zero. However we find that two chiral edge states, each localized at one of edges of the system defined on a cylinder coexist with the localized bulk states~\cite{sm}. The quasienergies of chiral edge states have a nontrivial flow under flux threading, which gives rise to a quantized pumped charge, as was previously observed in Ref.~\cite{titum}. The localized bulk states do not flow under threading of flux and hence do not contribute to the charge pumping. It was also shown in Ref.~\cite{titum} that if one of the edge modes is fully occupied, while the other remains unoccupied, then the net charge flowing across any bond on the occupied edge, per unit time remains quantized, and is equal to the winding number in the bulk when the system has been driven over many cycles. Here we show that the net charge pumped from the bulk to the edges when one quantum of flux is threaded through the cylinder also remains quantized.

Tuning $F$ away from $F=2$, for $\Omega/J=8.7$, increases the dispersion of the bulk bands which has a destabilizing effect on the AFAI phase. We find that the AFAI is stable between $1.9\leq F\leq 2.1$~\cite{sm}. Topological edge states have been observed in ultracold atoms by creating a programmable repulsive potential and releasing a localized Bose Einstein condensate near the edge using an optical tweezer. Subsequently, in the clean system, the wave packet propagates along the potential boundary, following its curvature, which is a characteristic for chiral edge states~\cite{braun}. Such chiral motion at the potential boundary should also be observable in the AFAI, while, in contrast, once the repulsive potential is switched off the initial wave packet should remain localized. For weak disorder, when the system is not in the AFAI phase, sufficiently high energy wave packets, within the first FBZ, will not remain localized, while for strong disorder when the system is in the AI phase, there should be no chiral motion at the edge.

\textit{Conclusion.}$-$We have studied localization properties and charge pumping in a disordered, circularly driven honeycomb lattice with a continuous driving protocol realized in the experiments~\cite{wintersperger,braun}. Within the scope of finite size numerics, we found that a new phase emerges at intermediate disorder strength, in which the time-averaged bulk states are fully localized while the system supports quantized charge pumping via edge states, when the system has been evolved over many driving cycles. This is the AFAI phase which was previously predicted in a simplified model~\cite{titum}, which is difficult to realize with ultracold atoms. We also show that the quantized charge pumping in the AFAI phase remains robust at intermediate disorder strength, in contrast to the CI phase.  Our approach will also allow us in the future to study the interplay of on-site interactions and strong disorder in the periodically driven system, which can lead to discovery of new phases in hitherto unexplored parameter regimes using Floquet-DMFT~\cite{tao,taoqin}.

\textit{Acknowledgements.}$-$This work was supported by the Deutsche Forschungsgemeinschaft (DFG, German Research Foundation) under Project No. 277974659 via Research Unit FOR 2414. J.-H. Z. acknowledges support from the NSFC under Grant No.12247103 and No.12175180, Shaanxi Fundamental Science Research Project for Mathematics and Physics under Grant No.\,22JSQ041 and No.\,22JSZ005, and the Youth Innovation Team of Shannxi Universities. M.A. also acknowledges support from the Deutsche Forschungsgemeinschaft (DFG) under Germany’s Excellence Strategy – EXC-2111 – 390814868. A.D. thanks Y. Xu for helpful discussions. The authors gratefully acknowledge the Gauss Centre for Supercomputing e.V. (www.gauss-centre.eu) for funding this project by providing computing time through the John von Neumann Institute for Computing (NIC) on the GCS Supercomputer JUWELS at J\"ulich Supercomputing Centre (JSC). Calculations for this research were also performed on the Goethe-NHR high performance computing cluster. The cluster is managed by the Center for Scientific Computing (CSC) of the Goethe University Frankfurt.


\begin{thebibliography}{99}
\bibitem{comment1} It can be easily verified that for two integers $n$ and $p$, if $u^{(n)}_{\bm{i}\alpha}$ satisfies $\sum_{\bm{j},m}\tilde{H}^{(n,m)}_{\bm{ij}}u_{\bm{j}\alpha}^{(m)} = \varepsilon_{\alpha}u_{\bm{i}\alpha}^{(n)}$ with the quasienergy $\varepsilon_{\alpha}$, then $u^{(n+p)}_{\bm{i}\alpha}$ also satisfies the same eigenvalue problem with quasienergy $\varepsilon_{\alpha}+p\Omega$
\bibitem{comment2} For different disorder realizations, the discontinuities appear at different $\theta$ values. In the numerical implementation, working with a fixed $\theta$ grid makes it impossible to capture the discontinuity for all disorder realizations. This further smears out the transition from the AFAI to the AI phase, and from the CI to AI phase.
\end{thebibliography}
\end{document}


\title{Supplementary materials: The anomalous Floquet Anderson insulator in a continuously driven optical lattice}

\author{Arijit Dutta}
\affiliation{Goethe-Universit\"at, Institut f\"ur Theoretische Physik, 
 60438 Frankfurt am Main, Germany}

\author{Efe Sen}
\affiliation{Goethe-Universit\"at, Institut f\"ur Theoretische Physik, 
 60438 Frankfurt am Main, Germany}
 
\author{Jun-Hui Zheng}
\affiliation{Shaanxi Key Laboratory for Theoretical Physics Frontiers, Institute of Modern Physics, Northwest University, Xi'an, 710127, China}
\affiliation{Peng Huanwu Center for Fundamental Theory, Xi'an 710127, China}

\author{Monika Aidelsburger}
\affiliation{Max-Planck-Institut f\"ur Quantenoptik, Hans-Kopfermann-Strasse 1, 85748 Garching, Germany}
\affiliation{Fakultät f\"ur Physik, Ludwig-Maximilians-Universität, Schellingstrasse 4, 80799 M\"unchen, Germany}
\affiliation{Munich Center for Quantum Science and Technology (MCQST), Schellingstraße 4, 80799 M\"unchen, Germany}

\author{Walter Hofstetter}
\affiliation{Goethe-Universit\"at, Institut f\"ur Theoretische Physik, 
 60438 Frankfurt am Main, Germany}

\date{\today}

\maketitle
\section{Floquet theory}
Floquet's theorem states that a time periodic Hamiltonian $H(t)$ with period $\tau$ which satisfies $H(t+\tau)=H(t)$ admits solutions of the form
\begin{equation}
    \ket{\psi_{\alpha}(t)} = e^{-i\varepsilon_{\alpha}t}\ket{u_{\alpha}(t)}
\end{equation}
where $\varepsilon_{\alpha}$ are the time-independent quasienergies while the states $\ket{u_{\alpha}(t)}$ are periodic in time with period $\tau$, i.e., $\ket{u_{\alpha}(t+\tau)} = \ket{u_{\alpha}(t)}$.
$\ket{u_{\alpha}(t)}$ can be expanded in terms of Floquet harmonics $\ket{u_{\alpha}^{n}}$, where
\begin{subequations}\label{eq:eom}
\begin{align}
\ket{u_{\alpha}^{n}} &= \frac{1}{\tau}\int\limits_{0}^{\tau}\mathrm{d}t\,e^{in\Omega t}\ket{u_{\alpha}(t)}\label{eq:eom1}\\
    \ket{u_{\alpha}(t)} &= \lim_{N\rightarrow\infty}\sum\limits_{n=-N/2}^{N/2}e^{-in\Omega t}\ket{u_{\alpha}^{n}}\label{eq:eom2}
\end{align}
\end{subequations}
where $\Omega\equiv 2\pi/\tau$ is the driving frequency. In terms of these discrete frequency modes, the Schr\"odinger equation
\begin{subequations}
\begin{align}
    i\frac{\partial}{\partial t}\ket{\psi_{\alpha}(t)} &= H(t)\ket{\psi_{\alpha}(t)}\\
    \qty(\varepsilon_{\alpha}+i\frac{\partial}{\partial t})\ket{u_{\alpha}(t)} &= H(t)\ket{u_{\alpha}(t)}\label{eq:eom-u}
\end{align}
can be recast into
\begin{align}
    \sum_{m}\tilde{H}^{(n-m)}\ket{u_{\alpha}^{m}} = \varepsilon_{\alpha}\ket{u_{\alpha}^{n}}\label{eq:quasi}
\end{align}
where
\begin{align}
    \tilde{H}^{(n-m)} = \frac{1}{\tau}\int\limits_{0}^{\tau}\mathrm{d}t\,e^{i\qty(n-m)\Omega t}H(t)-\delta_{nm}m\Omega
\end{align}
\end{subequations}
is the Fourier transform of the Hamiltonian $H(t)$. By choosing a complete set of Wannier-basis states $\{\ket{\bm{i}}\}$, $\tilde{H}$ can be diagonalized to obtain the quasienergies $\varepsilon_{\alpha n}\equiv\varepsilon_{\alpha}+n\Omega$ where $\alpha$ is an index for labelling the quasienergies in the first Floquet Brillouin zone and the wavefunctions for Floquet harmonics $u_{i\alpha}^{(n)}\equiv\braket{\bm{i}}{u_{\alpha}^{n}}$.

\section{Floquet Green's functions}
Let $c^{\dagger}_{\bm{i}}$ and $c_{\bm{i}}$ be the fermionic field operators which create and annihilate a particle, respectively, at site $\bm{i}$ and time $t$. These obey the fermionic anticommutation relation $\{c_{\bm{i}},c^{\dagger}_{\bm{j}}\} = \delta_{\bm{ij}}$. The single-particle retarded Green's function in the site representation is defined as
\begin{equation}
    G^{R}_{\bm{i};\bm{j}}(t,t^{\prime}) = -i\theta(t-t^{\prime})\langle\{c_{\bm{i}}(t),c^{\dagger}_{\bm{j}}(t^{\prime})\}\rangle\label{eq:GRdef}
\end{equation}
where the operators are in the Heisenberg representation and the average is taken with respect to the many-body state at some initial time $t_0$ which we can set to zero. In the following we shall only consider a noninteracting periodically driven system. 

In the Schr\"odinger picture, the creation operators for the Floquet modes are defined as
\begin{equation}
    c^{\dagger}_{\alpha}(t) = \sum_{\bm{i}}\braket{\bm{i}}{u_{\alpha}(t)}c^{\dagger}_{i}\label{eq:mode}
\end{equation}
Hence, $c^{\dagger}_{\alpha}(t)\ket{0}=\ket{u_{\alpha}(t)}$ and the annihilation operators are obtained by Hermitian conjugation. As a result, they satisfy the equal-time commutation relation
\begin{equation}
    \{c_{\alpha}(t),c^{\dagger}_{\beta}(t)\} = \sum_{\bm{ij}}\braket{u_{\alpha}(t)}{\bm{i}}\braket{\bm{j}}{u_{\beta}(t)}\{c_{\bm{i}},c^{\dagger}_{\bm{j}}\}=\delta_{\alpha\beta}
\end{equation}
Eq.~\eqref{eq:mode} can be inverted to get
\begin{equation}
    c^{\dagger}_{\bm{i}} = \sum_{\alpha}\braket{u_{\alpha}(t)}{\bm{i}}c^{\dagger}_{\alpha}(t)
\end{equation}
where we have used $\sum_{\alpha}\ket{u_{\alpha}(t)}\bra{u_{\alpha}(t)} = \mathbb{1}$ as resolution of the identity operator. Using these the time-dependent noninteracting Hamiltonian can be expressed as
\begin{equation}
    H(t) = \sum_{\alpha\beta}h_{\alpha\beta}(t)c^{\dagger}_{\alpha}(t)c_{\beta}(t)\label{eq:quad}
\end{equation}
where $h_{\alpha\beta}(t) = \mel{u_{\alpha}(t)}{H(t)}{u_{\beta}(t)}$. In the Heisenberg picture, the time evolution of $c_{\alpha}(t)$ is governed by
\begin{subequations}
\begin{align}\label{eq:heisen-c}
    \frac{d c_{\alpha}(t)}{dt} &= i\qty[H(t),c_{\alpha}(t)]+\frac{\partial c_{\alpha}}{\partial t} = -i\varepsilon_{\alpha}c_{\alpha}
\end{align}
where we have used Eq.~\eqref{eq:quad} and Eq.~\eqref{eq:eom-u} to simplify the right hand side of Eq.~\eqref{eq:heisen-c}.

This can be solved to obtain
\begin{align}
    c_{\alpha}(t) &= e^{-i\varepsilon_{\alpha}t}c_{\alpha}(0)\\
c^{\dagger}_{\alpha}(t) &= e^{i\varepsilon_{\alpha}t}c^{\dagger}_{\alpha}(0)
\end{align}
Hence,
\begin{equation}
    \{c_{\alpha}(t),c^{\dagger}_{\alpha}(t^{\prime})\} = e^{-i\varepsilon_{\alpha}(t-t^{\prime})}\delta_{\alpha\alpha^{\prime}}\label{eq:eom-comm}
\end{equation}
So, in the Heisenberg picture,
\begin{align}
    c_{\bm{i}}(t) &= \sum_{\alpha}\braket{\bm{i}}{u_{\alpha}(t)}e^{-i\varepsilon_{\alpha}t} c_{\alpha}(0)\label{eq:eom-c}\\
    c^{\dagger}_{\bm{i}}(t) &= \sum_{\alpha}\braket{u_{\alpha}(t)}{\bm{i}}e^{i\varepsilon_{\alpha}t}c^{\dagger}_{\alpha}(0)\label{eq:eom-cdagger}
\end{align}
\end{subequations}
Using Eq. \eqref{eq:eom-c} and Eq. \eqref{eq:eom-cdagger}, Eq. \eqref{eq:GRdef} can be simplified as
\begin{subequations}
\begin{equation}
    G^{R}_{\bm{i};\bm{j}}(t,t^{\prime}) = -i\Theta(t-t^{\prime})\sum_{\alpha}u_{\bm{i}\alpha}(t)u_{\bm{j}\alpha}^{*}(t^{\prime})e^{-i\varepsilon_{\alpha}(t-t^{\prime})}
\end{equation}
where $u_{\bm{i}\alpha}(t)\equiv\braket{\bm{i}}{u_{\alpha}(t)}=\sum_{n}e^{-i n\Omega t}u^{(n)}_{\bm{i}\alpha}$. By introducing the average time $t_a\equiv\qty(t+t^{\prime})/2$ and the relative time $t_r\equiv t-t^{\prime}$, we can define the retarded Floquet Green's function is defined as \cite{neqdmft}
\begin{equation}
    G^R_{\bm{i},m;\bm{j},n}(\omega)\equiv\int\limits_{0}^{\tau}\frac{\mathrm{d}t_{a}}{\tau}\int\limits_{-\infty}^{\infty}\mathrm{d}t_{r}\,e^{i\qty(\omega+m\Omega)t-i\qty(\omega+n\Omega) t^{\prime}}G^{R}_{\bm{i};\bm{j}}(t,t^{\prime})
\end{equation}
which can be simplified using the integral representation of the Heaviside function,
\begin{equation}
    \Theta(t-t^{\prime}) = -\lim_{\eta\rightarrow 0^+}\int\limits_{-\infty}^{\infty}\frac{\mathrm{d}z}{2\pi i}\, \frac{e^{-iz\qty(t-t^{\prime})}}{z+i\eta}
\end{equation}
\begin{widetext}
\begin{align}
    G^R_{\bm{i},m;\bm{j},n}(\omega)
    &= -i\sum_{\alpha}\int\limits_{0}^{\tau}\frac{\mathrm{d}t_{a}}{\tau}e^{i\qty(m-n)\Omega t_a}\int\limits_{-\infty}^{\infty}\mathrm{d}t_{r}\,e^{i\qty(\omega+\qty(\frac{m+n}{2})\Omega-\varepsilon_{\alpha})t_r}\Theta(t_r)u_{\bm{i}\alpha}(t_a+\frac{t_r}{2})u_{\bm{j}\alpha}^{*}(t_a-\frac{t_r}{2})\\
     &=\frac{1}{2\pi}\lim_{\eta\rightarrow 0^+}\sum_{\alpha;p,q}u_{\bm{i}\alpha}^{(p)}\qty[u_{\bm{j}\alpha}^{(q)}]^{*}\int\limits_{0}^{\tau}\frac{\mathrm{d}t_{a}}{\tau}e^{i\qty(m-n-p+q)\Omega t_a}\int\limits_{-\infty}^{\infty}\mathrm{d}z\int\limits_{-\infty}^{\infty}\mathrm{d}t_{r}\,\frac{e^{i\qty(\omega+\qty(\frac{m+n-p-q}{2})\Omega-\varepsilon_{\alpha}-z)t_r}}{z+i\eta}\\
     &=\lim_{\eta\rightarrow 0^+}\sum_{\alpha;p,q}\frac{u_{\bm{i}\alpha}^{(p)}\qty[u_{\bm{j}\alpha}^{(q)}]^{*}\delta_{m-n+q,p}}{\omega+\qty(\frac{m+n-p-q}{2})\Omega-\varepsilon_{\alpha}+i\eta} = \lim_{\eta\rightarrow 0^+}\sum_{\alpha,q}\frac{u_{\bm{i}\alpha}^{(m-n+q)}\qty[u_{\bm{j}\alpha}^{(q)}]^{*}}{\omega+\qty(n-q)\Omega-\varepsilon_{\alpha}+i\eta}
\end{align}
\end{widetext}
Next, by casting the sum over the Floquet harmonics in terms of a new variable $s=q-n$, we can write,
\begin{equation}
    G^R_{\bm{i},m;\bm{j},n}(\omega)  = \lim_{\eta\rightarrow 0^{+}}\sum_{\alpha,s}\frac{u^{(m+s)}_{\bm{i}\alpha}\qty[u^{(n+s)}_{\bm{j}\alpha}]^{*}}{\omega-\varepsilon_{\alpha s}+i\eta}
\end{equation}
where $\varepsilon_{\alpha s}\equiv\varepsilon_\alpha+s\Omega$. The advanced Floquet Green's function $\bm{G}^{A}(\omega) = \qty[\bm{G}^{R}(\omega)]^{\dagger}$ where a boldface indicates that the function should be considered as a matrix in the discrete indices.
\end{subequations}

\begin{subequations}
Next, we consider a scenario where each site in the system is coupled to a fermionic bath at temperature $T$ with identical coupling strength. The system-bath coupling is assumed to be bilinear and the baths at all the sites are assumed to be identical~\cite{tao2}.  In this case the nonequilibrium Floquet Green's functions (NFGF) $\bm{\mathcal{G}}^R$, $\bm{\mathcal{G}}^A$ and $\bm{\mathcal{G}}^K$ can be obtained by solving the Dyson's equation~\cite{neqdmft}
\begin{equation}
\begin{pmatrix}
\bm{\mathcal{G}}^{R}&\bm{\mathcal{G}}^{K}\\
\bm{0}&\bm{\mathcal{G}}^{A}
\end{pmatrix}^{-1} =
\begin{pmatrix}
\qty[\bm{G}^{R}]^{-1}&2i\eta\bm{\tilde{F}}\\
\bm{0}&\qty[\bm{G}^{A}]^{-1}
\end{pmatrix} 
-
\begin{pmatrix}
-i\Gamma\mathbb{1}&-2i\Gamma \bm{F}\\
\bm{0}&i\Gamma\mathbb{1}
\end{pmatrix}
\end{equation}
where $\Gamma$ is the site and energy independent (in the wide-bandwith limit) damping rate due to the bath and $T$ is the temperature of the bath. $\bm{\tilde{F}}$ denotes the distribution function in the system, which depends on its initial state and
\begin{equation}
    \bm{F}(\omega)\equiv F_{m}(\omega)\delta_{mn}\delta_{\bm{i}\bm{j}}=\tanh\qty(\frac{\omega+m\Omega}{2T})\delta_{mn}\delta_{\bm{i}\bm{j}}
\end{equation}
 Here we have used boldface notation to denote matrices. In presence of a bath the $\eta\rightarrow 0^{+}$ limit can be safely executed and the NFGF can be expressed as
\begin{align}
    \mathcal{G}^{R/A}_{\bm{i},m;\bm{j},n}(\omega) = \sum_{\alpha,s}\frac{u^{(m+s)}_{\bm{i}\alpha}\qty[u^{(n+s)}_{\bm{j}\alpha}]^{*}}{\omega-\varepsilon_{\alpha s}\pm i\Gamma}\\
    \bm{\mathcal{G}}^{K}(\omega) = -2i\Gamma\bm{\mathcal{G}}^{R}(\omega)\bm{F}(\omega)\bm{\mathcal{G}}^{A}(\omega)
\end{align}
Another relevant NFGF is the \emph{lesser} Green's function $\bm{\mathcal{G}}^{<}$, given by
\begin{equation}
    \bm{\mathcal{G}}^{<}(\omega) = \frac{1}{2}\qty[\bm{\mathcal{G}}^{K}(\omega)-\bm{\mathcal{G}}^{R}(\omega)+\bm{\mathcal{G}}^{A}(\omega)]
\end{equation}
\end{subequations}

 Next, we find an expression for $\bm{\mathcal{G}}^{<}$, explicitly, in terms of the Floquet mode wavefunctions and the quasienergies.
\begin{widetext}
\begin{subequations}
\begin{align}
    &\mathcal{G}^{<}_{\bm{i},m;\bm{j};n}(\omega) = -i\Gamma\sum\limits_{\bm{l},p}\mathcal{G}^R_{\bm{i},m;\bm{l},p}(\omega)F_{p}(\omega)\mathcal{G}^{A}_{\bm{l},p;\bm{j},n}(\omega) - \frac{1}{2}\mathcal{G}^R_{\bm{i},m,\bm{j},n} + \frac{1}{2}\mathcal{G}^A_{\bm{i},m,\bm{j},n}\\
    &= -i\Gamma\sum\limits_{\substack{\bm{l},p,\\\alpha,s,\alpha^{\prime},s^{\prime}}}\qty(\frac{u^{(m+s)}_{\bm{i}\alpha}\qty[u^{(p+s)}_{\bm{l}\alpha}]^{*}}{\omega-\varepsilon_{\alpha s}+i\Gamma})F_{p}(\omega)\qty(\frac{u^{(p+s)}_{\bm{l}\alpha^{\prime}}\qty[u^{(n+s)}_{\bm{j}\alpha^{\prime}}]^*}{\omega-\varepsilon_{\alpha^{\prime}s^{\prime}}-i\Gamma})
    +i\sum\limits_{\alpha,s}u^{(m+s)}_{\bm{i}\alpha}\qty[u^{(n+s)}_{\bm{j}\alpha}]^{*}\mbox{Im}\qty[\frac{1}{\omega-\varepsilon_{\alpha s}-i\Gamma}]\\
    &= i\Gamma\sum\limits_{\alpha s}\frac{u^{(m+s)}_{\bm{i}\alpha}\qty[u^{(n+s)}_{\bm{j}\alpha}]^{*}}{\qty(\omega-\varepsilon_{\alpha s})^2+\Gamma^2}\qty[1-\sum\limits_{\bm{l},p}F_p(\omega)\abs{u^{(p+s)}_{\bm{l}\alpha}}^2]
    -i\Gamma\sum\limits_{\substack{\bm{l},p\\\alpha\neq\alpha^{\prime}\\s\neq s^{\prime}}}F_p(\omega)\frac{u^{(m+s)}_{\bm{i}\alpha}\qty[u^{(p+s)}_{\bm{l}\alpha}]^{*}u^{(p+s^{\prime})}_{\bm{l}\alpha^{\prime}}\qty[u^{(n+s^{\prime})}_{\bm{j}\alpha^{\prime}}]^{*}}{\qty(\omega+i\Gamma-\varepsilon_{\alpha s})\qty(\omega-i\Gamma-\varepsilon_{\alpha^{\prime}s^{\prime}})}\label{eq:Gless}
\end{align}
\end{subequations}
\end{widetext}

\section{Simplification in the $\Gamma\rightarrow 0$ limit}
The pumped charge $P$ is defined in terms of the difference $\delta Q(\theta)$ between total occupancies $Q_{\mathcal{U}}= \sum_{\bm{i}\in\mathcal{U}}Q_{\bm{i}}$ and $Q_{\mathcal{L}}= \sum_{\bm{i}\in\mathcal{L}}Q_{\bm{i}}$ in the upper and the lower halves of the cylinder shown in Fig. $4 (a)$ in the main text, respectively.
\begin{subequations}
\begin{equation}
P \equiv (\max\{\delta Q(\theta)\}-\min\{\delta Q(\theta)\})/2    
\end{equation}
The site occupancies $Q_{\bm{i}}$ can be expressed in terms of the \emph{lesser} Floquet Green's function derived in Eq.~\eqref{eq:Gless}, 
\begin{align}\label{eq:Qi:0}
Q_{\bm{i}} &= \sum\limits_{n=\frac{-N}{2}}^{\frac{N}{2}} 
\int\limits_{-\Omega/2}^{\Omega/2}\frac{\mathrm{d}\omega}{2\pi}\,
\lim_{\Gamma\rightarrow0^{+}} \mbox{Im}\qty[G^{<}_{\bm{i},n;\bm{i},n}(\omega;\theta)]\\
&=\frac{1}{2}\sum\limits_{n,\alpha}\abs{u^{(n)}_{\bm{i}\alpha}}^2\qty[1-\sum\limits_{\bm{l},p}F_p(\omega)\abs{u^{(p)}_{\bm{l}\alpha}}^2]-\Lambda_{\bm{i}}
\end{align}
where $\Lambda_{\bm{i}}$ is given by
\begin{widetext}
\begin{align}\label{eq:Qi:1}
\Lambda_{\bm{i}} &=\sum\limits_{\substack{n,p,\bm{l}\\\alpha\neq\alpha^{\prime}\\s\neq s^{\prime}}}\int\limits_{-\Omega/2}^{\Omega/2}\frac{\mathrm{d}\omega}{2\pi}\,F_p(\omega)\lim_{\Gamma\rightarrow0^{+}}\mbox{Re}\qty[\frac{\Gamma u^{(n+s)}_{\bm{i}\alpha}\qty[u^{(p+s)}_{\bm{l}\alpha}]^{*}u^{(p+s^{\prime})}_{\bm{l}\alpha^{\prime}}\qty[u^{(n+s^{\prime})}_{\bm{i}\alpha^{\prime}}]^{*}}{\qty(\omega+i\Gamma-\varepsilon_{\alpha s})\qty(\omega-i\Gamma-\varepsilon_{\alpha^{\prime}s^{\prime}})}]\\
&=\sum\limits_{\substack{n,p,\bm{l}\\\alpha\neq\alpha^{\prime}\\s\neq s^{\prime}}}\int\limits_{-\Omega/2}^{\Omega/2}\frac{\mathrm{d}\omega}{2\pi}\,F_p(\omega)\lim_{\Gamma\rightarrow0^{+}}\qty[\mbox{Re}\qty(\frac{u^{(n+s)}_{\bm{i}\alpha}\qty[u^{(p+s)}_{\bm{l}\alpha}]^{*}u^{(p+s^{\prime})}_{\bm{l}\alpha^{\prime}}\qty[u^{(n+s^{\prime})}_{\bm{i}\alpha^{\prime}}]^{*}}{\varepsilon_{\alpha s}-\varepsilon_{\alpha^{\prime}s^{\prime}}-2i\Gamma})\qty(\frac{\Gamma\qty(\omega-\varepsilon_{\alpha s})}{\qty(\omega-\varepsilon_{\alpha s})^2+\Gamma^2}-\frac{\Gamma\qty(\omega-\varepsilon_{\alpha^{\prime} s^{\prime}})}{\qty(\omega-\varepsilon_{\alpha^{\prime} s^{\prime}})^2+\Gamma^2})]\nonumber\\
&-\sum\limits_{\substack{n,p,\bm{l}\\\alpha\neq\alpha^{\prime}\\s\neq s^{\prime}}}\int\limits_{-\Omega/2}^{\Omega/2}\frac{\mathrm{d}\omega}{2\pi}\,F_p(\omega)\lim_{\Gamma\rightarrow0^{+}}\qty[\mbox{Im}\qty(\frac{u^{(n+s)}_{\bm{i}\alpha}\qty[u^{(p+s)}_{\bm{l}\alpha}]^{*}u^{(p+s^{\prime})}_{\bm{l}\alpha^{\prime}}\qty[u^{(n+s^{\prime})}_{\bm{i}\alpha^{\prime}}]^{*}}{\varepsilon_{\alpha s}-\varepsilon_{\alpha^{\prime}s^{\prime}}-2i\Gamma})\qty(\frac{\Gamma^2}{\qty(\omega-\varepsilon_{\alpha s})^2+\Gamma^2}-\frac{\Gamma^2}{\qty(\omega-\varepsilon_{\alpha^{\prime} s^{\prime}})^2+\Gamma^2})]\\
&= 0
\end{align}
\end{widetext}
\end{subequations}
Hence, Eq.~\eqref{eq:Qi:0} simplifies to
\begin{align}
&Q_{\bm{i}} = \sum\limits_{\alpha}\qty(\sum\limits_{n}\abs{u^{(n)}_{\bm{i}\alpha}}^2)\qty(\sum\limits_{\bm{l},p}f(\varepsilon_{\alpha}+p\Omega)\abs{u^{(p)}_{\bm{l}\alpha}}^2)
\end{align}
where $f(\omega)\equiv\qty(1+\exp\qty(\omega/T))^{-1}$.

\begin{figure*}
    \centering
    \includegraphics[trim={0 1.25cm 0 0.2cm}, clip=True, width=0.75\linewidth]{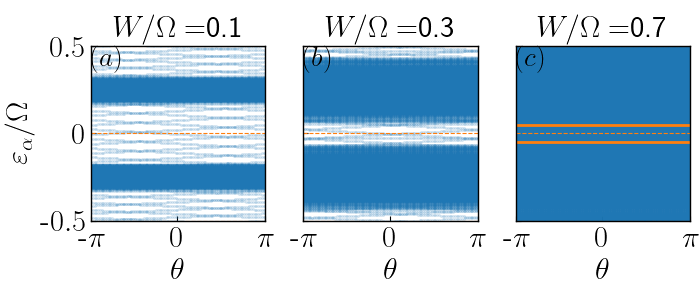}
    \includegraphics[width=0.75\linewidth]{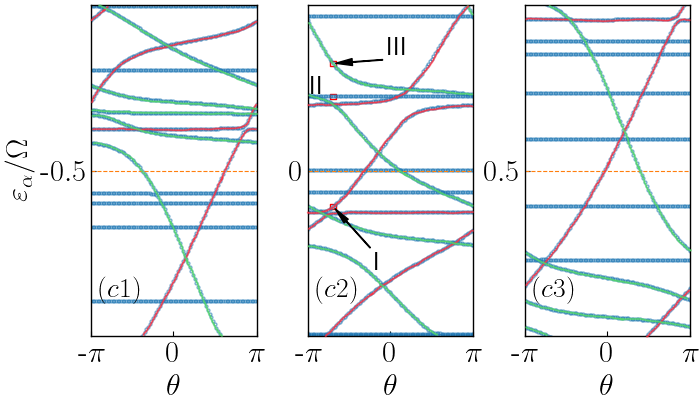}
    \includegraphics[trim={0 0.4cm 0 0.5cm}, clip=True, width=0.75\linewidth]{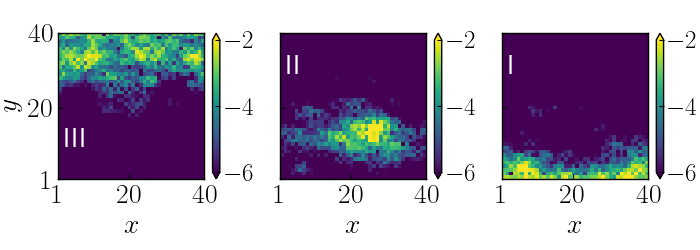}
    \caption{(Top row): Variation of the quasienergy spectrum with changing flux $\theta$ for a single disorder realization at the disorder strengths $W/\Omega = 0.1\,(a),\, 0.3\,(b)$ and $0.7\,(c)$. The system is in the AFAI phase at $W/\Omega = 0.7$. (Middle row): The resolved spectrum near the quasienergies $-\Omega/2\,(c1)$, $0\,(c2)$ and $\Omega/2\,(c3)$ for $W/\Omega = 0.7$. As theta is varied from $-\pi$ to $\pi$, the edge mode shown in red (green) winds around the FBZ in (anti-clockwise) clockwise sense. (Bottom row): $\overline{\abs{u_{\bm{i}\alpha}}^{2}}$, where $\alpha$ labels the corresponding quasienergies $\mbox{I},\,\mbox{II}$ and $\mbox{III}$ in $(c2)$ has been shown in log scale. The system size is $40\times 40$.}
    \label{fig:flow}
\end{figure*}
\begin{figure*}[t]
    \centering
    \includegraphics[width=\linewidth]{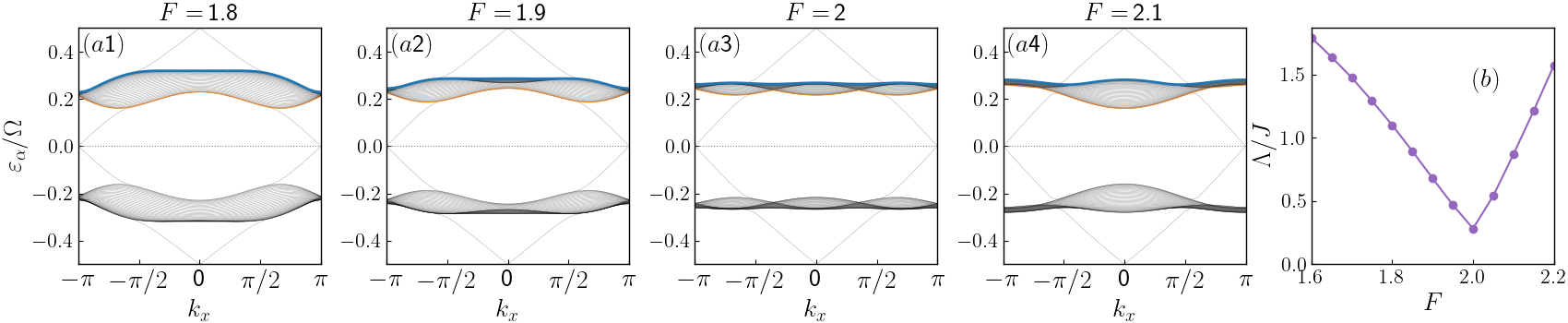}
    \includegraphics[width=\linewidth]{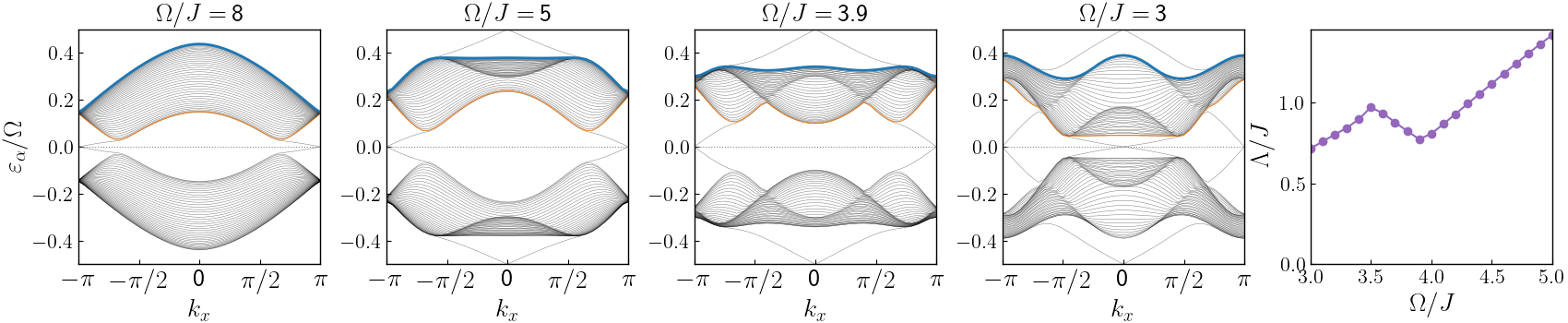}
    \caption{Top row: Effect of changing the parameter $F$ on the band structure of the clean system. An energy scale $\Lambda$ can be associated with the dispersion of the bulk bands (see Eq.~\ref{eq:Lambda}) which has a minimum at $F=2$ for $\Omega/J = 8.7$. Bottom row: Effect of changing $\Omega/J$ on the band structure of the clean system for $F=0.813$. For $3.5\leq\Omega/J < 4.5$, the values attained by $\Lambda/J$ are comparable to those attained by it around $F=1.9$ and $2.1$ and $\Omega/J = 8.7$.}
    \label{fig:bsFvar}
\end{figure*}
\section{Spectral flow in the disordered system}
\begin{figure}[b]
    \centering
    \includegraphics[width=\linewidth]{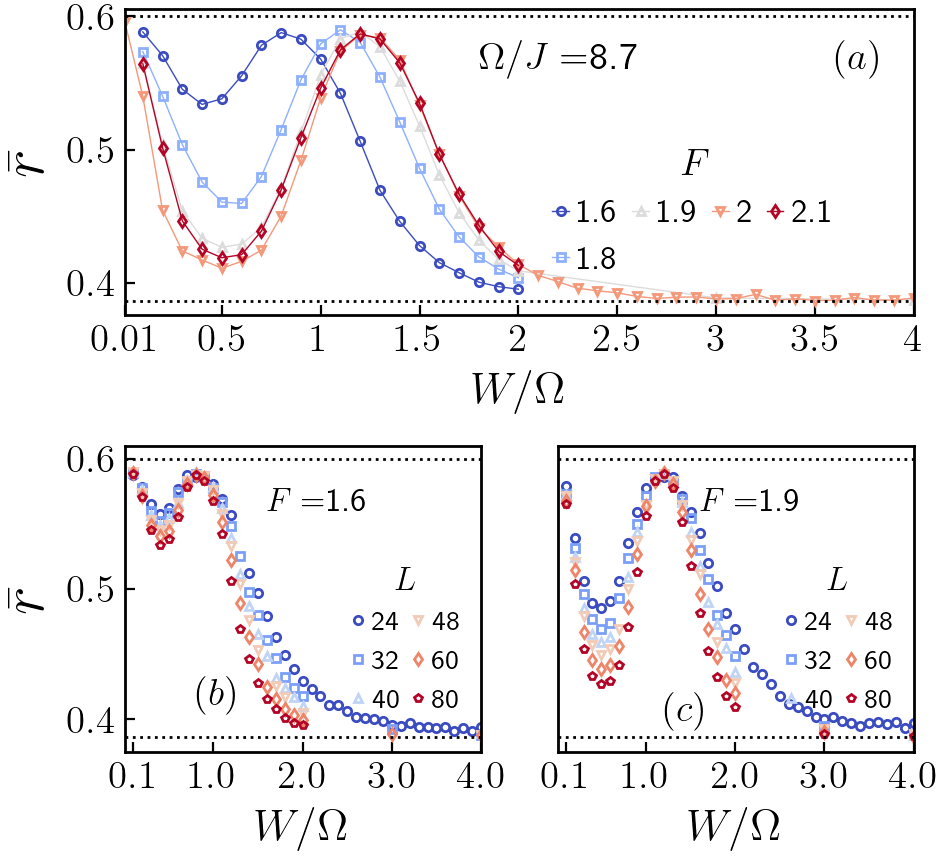}
    \caption{(a) The average level spacing ratio $\bar{r}$ with changing $W/\Omega$ for different values of $F$ for the system size $80\times 80$. Tuning $F$ away from $F=2$ shows a tendency towards delocalization. The finite-size scaling of $\bar{r}$ at $F = 1.6$ (b) and $F = 1.9$ (c) shows that increasing system size enhances the dip at $W/\Omega=0.5$.}
    \label{fig:lsrFnL}
\end{figure}
For the system defined on a cylinder, the quasienergy spectrum shows a remarkable dependence on the flux $\theta$ threaded through the cylinder. We show the behaviour of the quasienergies in the first FBZ, for $\theta\in\qty[-\pi,\pi]$, in Fig.~\ref{fig:flow} for $W/\Omega =\, (a)\,0.1,\, (b)\,0.3$ and $(c)\,0.7$. For $W/\Omega = 0.1$, and $0.3$, we notice that the there are two distinct bands separated by a gap which is traversed by the edge modes with changing $\theta$. The bands arise because the bulk states do not flow with changing $\theta$. At $W/\Omega = 0.7$, we find no clear demarcation between the bulk bands and the edge modes in the original resolution. To resolve the spectrum better, horizontal cuts of width $0.005$ are selected around $\varepsilon/\Omega = -0.5,\, 0$ and $0.5$ and the quasienergies in these intervals are plotted against $\theta$ in Fig.~\ref{fig:flow} $(c1)$, $(c2)$ and $(c3)$, respectively. We find that the quasienergies show three kinds of characteristic behaviour: (i) The quasienergies for bulk modes (shown in blue) do not disperse with changing $\theta$. This is confirmed by plotting, in Fig.~\ref{fig:flow} ($\mbox{II}$), the time averaged amplitude of the wavefunction for a non-dispersing mode at point $\mbox{II}$ (indicated by a red square) in Fig.~\ref{fig:flow} (c2). Recall that, for $0.4\leq W/\Omega\leq 0.7$, the bulk states are completely localised for sufficiently large $L$. In this regime the quasienergies of the bulk states do not depend on the value of the threaded flux and they trace constant paths as $\theta$ is varied over one period. (ii) As $\theta$ is varied from $-\pi$ to $\pi$, a mode (indicated by red) winds around the FBZ in an anti-clockwise sense. This mode is found to be localised in the lower edge of the cylinder [Fig.~\ref{fig:flow} ($\mbox{I}$)]. (iii) Another mode localized at the upper edge of the cylinder [Fig.~\ref{fig:flow} ($\mbox{III}$)] is also discernible which winds around the FBZ in a clockwise sense as $\theta$ varies from $-\pi$ to $\pi$.

\section{Effect of band dispersion on the localization properties of the bulk}
\begin{figure*}
    \centering
    \includegraphics[trim={0 0cm 0 0cm}, clip=True, width=0.45\linewidth]{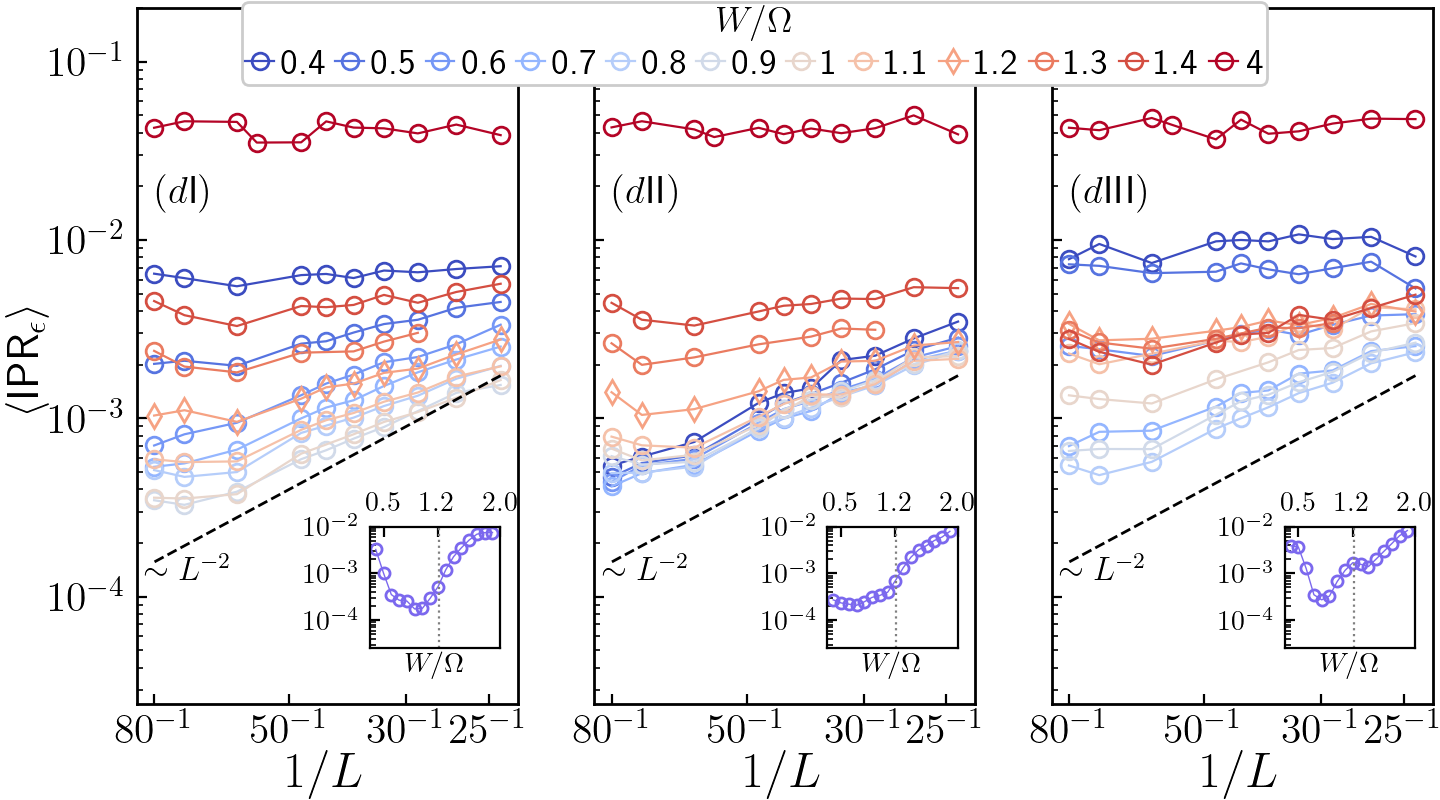}
    \includegraphics[trim={0 0cm 0 0cm}, clip=True, width=0.45\linewidth]{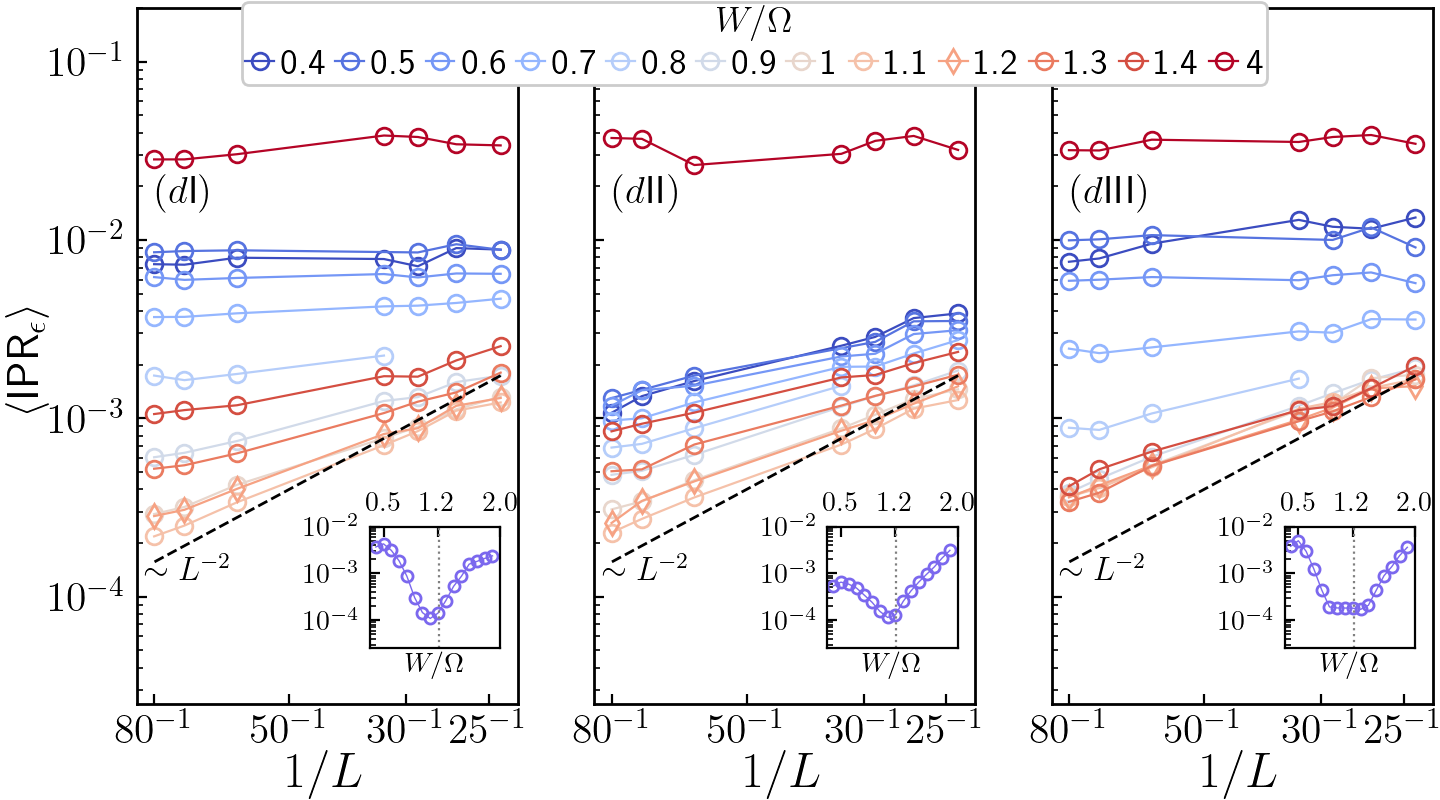}
    \includegraphics[trim={0 0cm 0 0cm}, clip=True, width=0.45\linewidth]{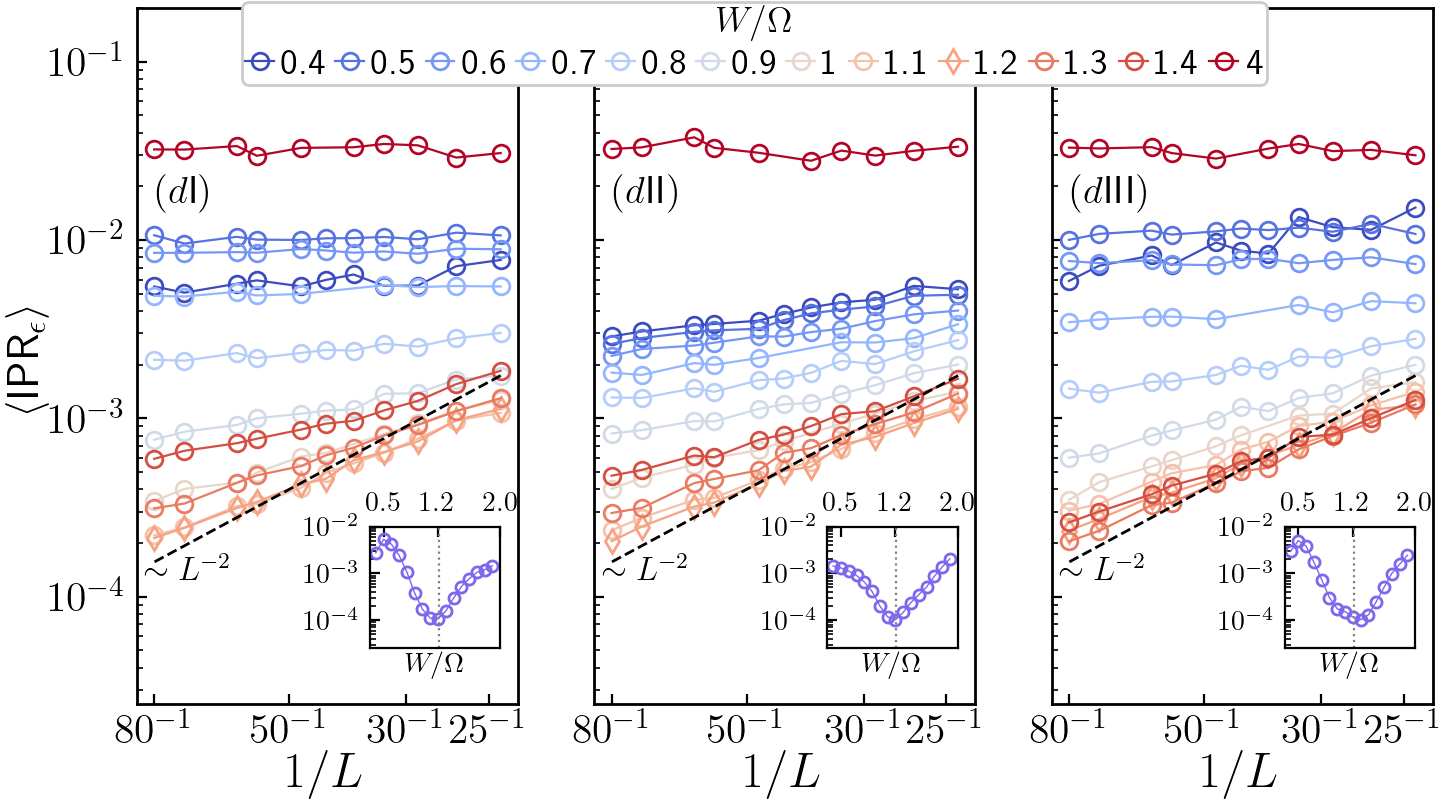}
    \includegraphics[trim={0 0cm 0 0cm}, clip=True, width=0.45\linewidth]{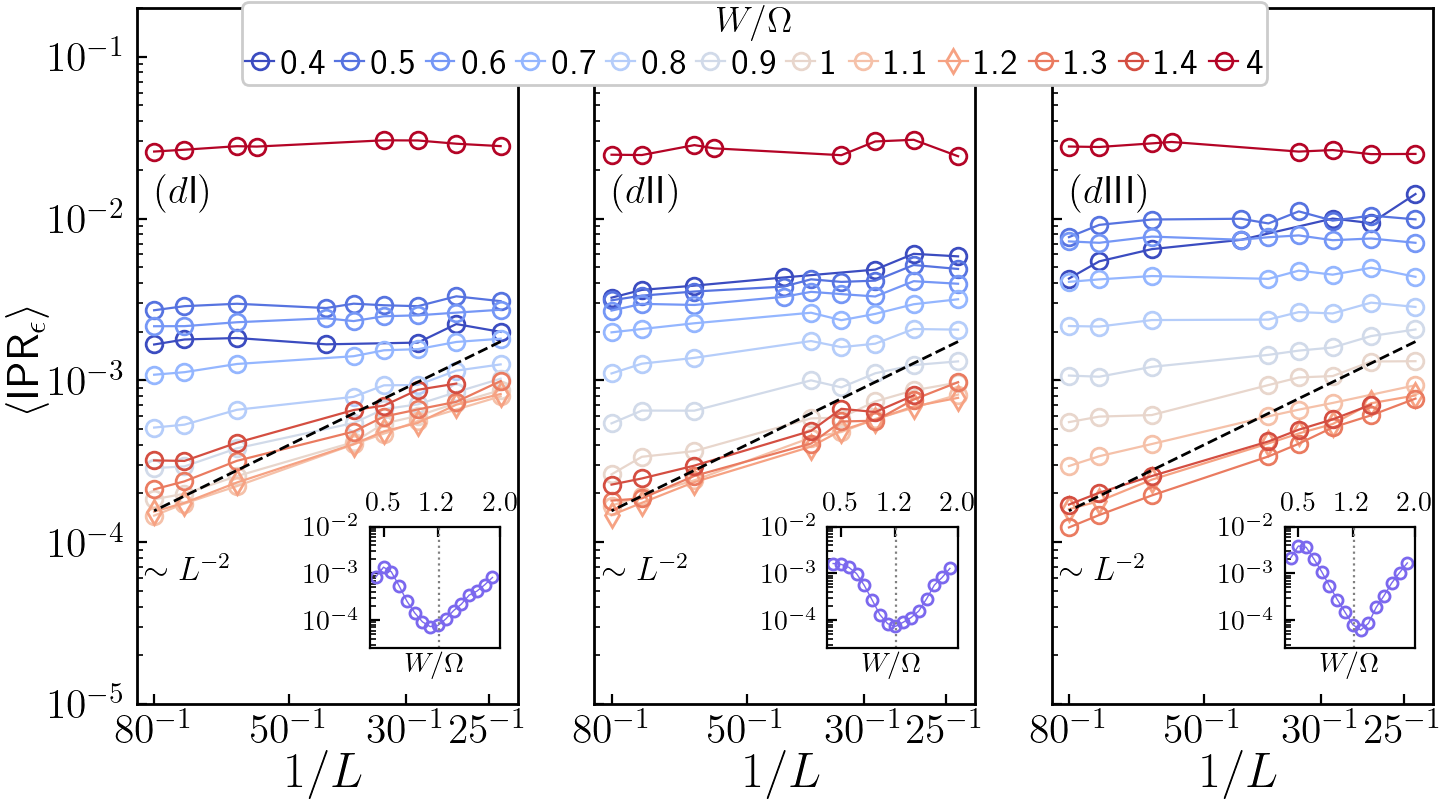}
    \caption{Size dependence of the disorder averaged IPR at three different quasienergy slices in the first Floquet Brillouin zone at $\Omega/J=8.7$ and different values of $W/\Omega$. $(a\mbox{I})-(a\mbox{III})$ corresponds to $F=1.6$, while $(b\mbox{I})-(b\mbox{III})$, $(c\mbox{I})-(c\mbox{III})$ and $(d\mbox{I})-(d\mbox{III})$ correspond to $F = 1.8,\, 1.9$ and $2.1$ respectively.}
    \label{fig:ipr}
\end{figure*}

In order to realize the AFAI phase all the time-averaged bulk states must be localized. The localization of the bulk states is dependent on the dispersion of the bulk bands, which itself depends on the driving frequency $\Omega$ and the parameter $F$. An energy scale $\Lambda$ can be associated with the dispersion of the bulk bands by evaluating the difference between the maximum and the minimum width of the band as a function of $k_x$, as shown in Fig.~\ref{fig:bsFvar}.
\begin{align}
    \Lambda(k_x) &\equiv \varepsilon_{\alpha_{+}}(k_x)-\varepsilon_{\alpha_{-}}(k_x)\nonumber\\
    \Lambda &\equiv \max_{k_x}\qty[\Lambda_{k_x}]-\min_{k_x}\qty[\Lambda_{k_x}]\label{eq:Lambda}
\end{align}
where $\alpha_{+(-)}$ is the index of the largest (smallest) quasienergy in a given bulk band (indicated by blue (orange) in Fig.~\ref{fig:bsFvar}). If the bulk band is perfectly non-dispersive, then $\Lambda = 0$ by construction. For a given $\Omega/J$, $\Lambda/J$ can be tuned by varying $F$. For $\Omega/J = 8.7$, within the AFTI phase, we find that $\Lambda/J$ has a cusp at $F=2$ [Fig.~\ref{fig:bsFvar} (top)]. In the main text we showed that the AFAI phase is stabilized at $\Omega/J = 8.7$ and $F=2$ for $0.4\leq W/\Omega\leq 0.7$ for system sizes closes to $80\times 80$. Here, we discuss the stability of the AFAI with changing $\Lambda$ within our finite-sized calculation.

We first examine the variation of the average level spacing ratio ($\bar{r}$) with changing $F$ at $\Omega/J=8.7$ for the largest accessible size ($80\times 80$). It can be seen in Fig.~\ref{fig:lsrFnL}(a) that away from $F=2$ the dip in $\bar{r}$ at $W/\Omega = 0.5$ diminishes, indicating that the bulk becomes less localized. For $F=1.6$ [Fig.~\ref{fig:lsrFnL}(b)] and $F=1.9$ [Fig.~\ref{fig:lsrFnL}(c)], finite-size scaling shows that the dip grows with increasing $L$, indicating a tendency towards localization with increasing system size. However, for the AFAI to be realized the bulk states at all quasienergies must be localized. This cannot be resolved by looking at $\bar{r}$. 

To resolve whether the time-averaged bulk states at all quasienergies are localized, we evaluate the inverse participation ratio (IPR) at three different quasienergy slices in the first Floquet Brillouin zone for different system sizes. The results are shown in Fig.~\ref{fig:ipr}. We find that even though the average LSR calculation indicates a tendency towards localization with increasing system size for $1.6\leq F\leq 2.1$, the size dependence of IPR shows that for $F < 1.9$ the bulk states at all the three quasienergy slices are not localized, even for the largest system size which could be accessed.